\documentclass[longauth]{aa}

\usepackage{graphicx}
\usepackage{natbib}
\usepackage{scalerel}
\usepackage{comment}

\usepackage[table]{xcolor}
\usepackage{multirow}
\usepackage{txfonts}
\usepackage[pdfencoding=auto,psdextra]{hyperref}
\hypersetup{
    colorlinks=true,
    linkcolor=blue,
    filecolor=magenta,      
    urlcolor=blue,
    citecolor=blue
}
\makeatletter
\renewcommand*\aa@pageof{, page \thepage{} of \pageref*{LastPage}}
\makeatother

\usepackage[utf8]{inputenc}

\usepackage[switch, modulo]{lineno}
\AtBeginDocument{\let\linenumbers\relax}
\usepackage{euclid}
\usepackage[nameinlink,capitalise]{cleveref}
\crefname{section}{Sect.}{Sects.}
\Crefname{section}{Section}{Sections}
\crefname{figure}{Fig.}{Figs.}
\Crefname{figure}{Figure}{Figures}
\crefname{equation}{Eq.}{Eqs.}
\Crefname{equation}{Equation}{Equations}
\crefname{table}{Table}{Tables}
\crefname{appendix}{Appendix}{Appendices}
\usepackage{xspace}
\providecommand{\Gaia}{\textit{Gaia}\xspace}
\defcitealias{Q1-TP002}{Q1VIS}
\newcommand{\fnurl}[2]{#1\footnote{\href{#2}{\url{#2}}}}
\usepackage[nolist,nohyperlinks]{acronym}

\begin{document}

%
%
\acrodef{AA}{azimuth or $\alpha$ angle}
\acrodef{ADC}{analogue-to-digital converter}
\acrodef{ADU}{analogue-to-digital unit}
\acrodef{AOI}{angle of incidence}
\acrodef{ASIC}{application specific integrated circuit}
\acrodef{BFE}{brighter-fatter effect}
\acrodef{CaLA}{camera-lens assembly}
\acrodef{CCD}{charge-coupled device}
\acrodef{CoLA}{corrector-lens assembly}
\acrodef{CDS}{Correlated Double Sampling}
\acrodef{CFC}{cryo-flex cable}
\acrodef{CFHT}{Canada-France-Hawaii Telescope}
\acrodef{CFRP}{carbon-fibre reinforced plastic}
\acrodef{CGH}{computer-generated hologram}
\acrodef{CME}{coronal mass ejection}
\acrodef{CNES}{Centre National d'Etude Spacial}
\acrodef{CPPM}{Centre de Physique des Particules de Marseille}
\acrodef{CPU}{central processingunit}
\acrodef{CR}{cosmic ray}
\acrodef{CTE}{coefficient of thermal expansion}
\acrodef{CME}{coronal mass ejection}
\acrodef{CTI}{charge-transfer inefficiency}
\acrodef{DCU}{Detector Control Unit}
\acrodef{DES}{Dark Energy Survey}
\acrodef{DPU}{Data Processing Unit}
\acrodef{DS}{Detector System}
\acrodef{EDS}{Euclid Deep Survey}
\acrodef{EE}{encircled energy}
\acrodef{EPER}{extended pixel-edge response}
\acrodef{ESA}{European Space Agency}
\acrodef{ESP}{Emission of Solar Protons}
\acrodef{ECSS}{European Cooperation for Space Standardization}
\acrodef{EWS}{Euclid Wide Survey}
\acrodef{EDS}{Euclid Deep Survey}
\acrodef{FDIR}{Fault Detection, Isolation and Recovery}
\acrodef{FGS}{fine guidance sensor}
\acrodef{FOM}{figure of merit}
\acrodef{FOV}{field of view}
\acrodef{FPA}{focal-plane array}
\acrodef{FPR}{false positive rate}
\acrodef{FWA}{filter-wheel assembly}
\acrodef{FWC}{full-well capacity}
\acrodef{FWHM}{full width at half maximum}
\acrodef{GOES}{Geostationary Operational Environmental Satellites}
\acrodef{GCR}{Galactic cosmic ray}
\acrodef{GWA}{grism-wheel assembly}
\acrodef{H2RG}{HAWAII-2RG}
\acrodef{HST}{\textit{Hubble} Space Telescope}
\acrodef{IP2I}{Institut de Physique des 2 Infinis de Lyon}
\acrodef{ICRS}{International Celestial Reference System}
\acrodef{JWST}{{\em James Webb} Space Telescope}
\acrodef{IAD}{ion-assisted deposition}
\acrodef{ICU}{Instrument Control Unit}
\acrodef{IPC}{inter-pixel capacitance}
\acrodef{ISES}{International Space Environmental Services}
\acrodef{JWST}{\textit{James Webb} Space Telescope}
\acrodef{LAM}{Laboratoire d'Astrophysique de Marseille}
\acrodef{LED}{light-emitting diode}
\acrodef{LSB}{low surface brightness}
\acrodef{MACC}{multiple accumulated}
\acrodef{MEF}{multi-extension \ac{FITS}}
\acrodef{FITS}{Flexible Image Transport System}
\acrodef{MER PF}{MER processing function}
\acrodef{MLI}{multi-layer insulation}
\acrodef{MMU}{Mass Memory Unit}
\acrodef{MPE}{Max-Planck-Institut für extraterrestrische Physik}
\acrodef{MPIA}{Max-Planck-Institut für Astronomie}
\acrodef{NA}{numerical aperture}
\acrodef{NASA}{National Aeronautic and Space Administration}
\acrodef{NIEL}{non-ionising energy loss}
\acrodef{JPL}{NASA Jet Propulsion Laboratory}
\acrodef{MZ-CGH}{multi-zonal computer-generated hologram}
\acrodef{NI-CU}{NISP calibration unit}
\acrodef{NI-OA}{near-infrared optical assembly}
\acrodef{NI-GWA}{NISP Grism Wheel Assembly}
\acrodef{NIR}{near-infrared}
\acrodef{NISP}{Near-Infrared Spectrometer and Photometer}
\acrodef{NOAA}{National Oceanic and Atmospheric Administration}
\acrodef{PARMS}{plasma-assisted reactive magnetron sputtering}
\acrodef{PLM}{payload module}
\acrodef{PRNU}{pixel-response non-uniformity}
\acrodef{PTC}{photon transfer curve}
\acrodef{PTFE}{polytetrafluoroethylene}
\acrodef{PV}{performance verification}
\acrodef{PWM}{pulse-width modulation}
\acrodef{PSF}{point-spread function}
\acrodef{QE}{quantum efficiency}
\acrodef{QF}{quality factor}
\acrodef{ROE}{readout electronic block unit}
\acrodef{ROI}{region of interest}
\acrodef{ROIC}{readout-integrated circuit}
\acrodef{ROS}{reference observing sequence}
\acrodef{SAA}{Solar aspect angle}
\acrodef{SCA}{sensor chip array}
\acrodef{SCE}{sensor chip electronic}
\acrodef{SCS}{sensor chip system}
\acrodef{SGS}{Science Ground Segment}
\acrodef{SGPS}{Solar and Galactic Proton Sensor}
\acrodef{SHS}{Shack-Hartmann sensor}
\acrodef{SNR}[S/N]{signal-to-noise ratio}
\acrodef{SED}{spectral energy distribution}
\acrodef{SiC}{silicon carbide}
\acrodef{SEP}{Solar energetic particle}
\acrodef{SSN}{Sunspot number}
\acrodef{STIX}{Spectrometer/Telescope for Imaging X-rays}
\acrodef{SolO}{Solar Orbiter}
\acrodef{TP}{trap pumping}
\acrodef{TPR}{true positive rate}
\acrodef{UTR}{up-the-ramp}
\acrodef{SVM}{service module}
\acrodef{VIS}{visible imager}
\acrodef{WD}{white dwarf}
\acrodef{WCS}{world coordinate system}
\acrodef{WFE}{wavefront error}
\acrodef{ZP}{zero point}
\acrodef{DPDD}{Data Product Definition Document}
\acrodef{OTF}{on-the-fly}
\acrodef{RR}{regression reprocessing}
\acrodef{EUDF}{Euclid ultra-deep field}
\acrodef{PF}{processing function}
\acrodef{PA}{position angle}
\acrodef{RMS}{root mean square} 
\acrodef{PE}{processing element} 
\acrodef{NL}{non-linearity} 
\acrodef{QSO}{quasi-stellar object}
\acrodef{Q1}{Quick Release 1}
\acrodef{Q2}{Quick Release 2}
\acrodef{DR1}{Data Release 1}
\acrodef{DR2}{Data Release 2}
\acrodef{SOC}{science operations centre}
\acrodef{EDFN}[EDF-N]{Euclid Deep Field North}
\acrodef{MER}{MERging}
\acrodef{HSC}{Hyper Suprime-Cam}
\acrodef{IRAC}{Infrared Array Camera}
\acrodef{PRIMER}{Public Release IMaging for Extragalactic Research}
\acrodef{NIRCam}{Near Infrared Camera}
\acrodef{EDF-N}[EDF-N]{Euclid Deep Field North}
\acrodef{EDF-S}[EDF-S]{Euclid Deep Field South}
\acrodef{EDF-F}[EDF-F]{Euclid Deep Field Fornax}
\acrodef{MER}{MERge}
\acrodef{HSC}{Hyper Suprime-Cam}
\acrodef{IRAC}{Infrared Array Camera}
\acrodef{PRIMER}{Public Release IMaging for Extragalactic Research}
\acrodef{NIRCam}{Near Infrared Camera}
\acrodef{DQC}{Data quality control}
\acrodef{HKTM}{housekeeping telemetry} 
\acrodef{RSU}{readout shutter unit}

\title{Euclid Data Release 1 (DR1)}
\subtitle{VIS processing and data products}   
\maxdeadcycles=500 


\newcommand{\orcid}[1]{} 
\author{Euclid Collaboration: H.~J.~McCracken\orcid{0000-0002-9489-7765}\thanks{\email{henry.mc\_cracken@sorbonne-universite.fr}}\inst{\ref{aff1}}
\and K.~Benson\inst{\ref{aff2}}
\and M.~Cropper\orcid{0000-0003-4571-9468}\inst{\ref{aff2}}
\and C.~Dolding\orcid{0009-0003-7199-6108}\inst{\ref{aff2}}
\and T.~Flanet\inst{\ref{aff1}}
\and C.~Grenet\inst{\ref{aff1}}
\and O.~Herent\inst{\ref{aff1}}
\and P.~Hudelot\inst{\ref{aff1}}
\and C.~Laigle\orcid{0009-0008-5926-818X}\inst{\ref{aff1}}
\and G.~Leroy\orcid{0009-0004-2523-4425}\inst{\ref{aff3},\ref{aff4}}
\and P.~Liebing\orcid{0009-0005-8043-7150}\inst{\ref{aff2}}
\and R.~J.~Massey\orcid{0000-0002-6085-3780}\inst{\ref{aff4}}
\and L.~Miller\orcid{0000-0002-3376-6200}\inst{\ref{aff5}}
\and S.~Mottet\inst{\ref{aff1}}
\and R.~Nakajima\orcid{0009-0009-1213-7040}\inst{\ref{aff6}}
\and H.~N.~Nguyen-Kim\inst{\ref{aff1}}
\and M.~J.~Page\orcid{0000-0002-6689-6271}\inst{\ref{aff2}}
\and J.~Skottfelt\orcid{0000-0003-1310-8283}\inst{\ref{aff7}}
\and F.~Soldano\inst{\ref{aff1}}
\and M.~von~Wietersheim-Kramsta\orcid{0000-0003-4986-5091}\inst{\ref{aff4},\ref{aff3}}
\and E.~Vilenius\orcid{0000-0002-6184-7681}\inst{\ref{aff2}}
\and H.~Aussel\orcid{0000-0002-1371-5705}\inst{\ref{aff8}}
\and P.~Casenove\orcid{0009-0006-6736-1670}\inst{\ref{aff9}}
\and G.~Libet\inst{\ref{aff9}}
\and K.~Paterson\orcid{0000-0001-8340-3486}\inst{\ref{aff10}}
\and M.~Ruffenach\orcid{0000-0001-8772-7330}\inst{\ref{aff9}}
\and M.~Schirmer\orcid{0000-0003-2568-9994}\inst{\ref{aff10}}
\and L.~C.~Smith\orcid{0000-0002-3259-2771}\inst{\ref{aff11}}
\and A.~Zacchei\orcid{0000-0003-0396-1192}\inst{\ref{aff12},\ref{aff13}}
\and B.~Altieri\orcid{0000-0003-3936-0284}\inst{\ref{aff14}}
\and A.~Amara\inst{\ref{aff15}}
\and S.~Andreon\orcid{0000-0002-2041-8784}\inst{\ref{aff16}}
\and N.~Auricchio\orcid{0000-0003-4444-8651}\inst{\ref{aff17}}
\and S.~Awan\inst{\ref{aff2}}
\and R.~Azzollini\orcid{0000-0002-0438-0886}\inst{\ref{aff2}}
\and C.~Baccigalupi\orcid{0000-0002-8211-1630}\inst{\ref{aff13},\ref{aff12},\ref{aff18},\ref{aff19}}
\and M.~Baldi\orcid{0000-0003-4145-1943}\inst{\ref{aff20},\ref{aff17},\ref{aff21}}
\and S.~Bardelli\orcid{0000-0002-8900-0298}\inst{\ref{aff17}}
\and A.~Basset\orcid{0009-0001-5490-3139}\inst{\ref{aff9}}
\and A.~N.~Belikov\inst{\ref{aff22},\ref{aff23}}
\and R.~Bender\orcid{0000-0001-7179-0626}\inst{\ref{aff24},\ref{aff25}}
\and A.~Biviano\orcid{0000-0002-0857-0732}\inst{\ref{aff12},\ref{aff13}}
\and M.~Bolzonella\orcid{0000-0003-3278-4607}\inst{\ref{aff17}}
\and E.~Branchini\orcid{0000-0002-0808-6908}\inst{\ref{aff26},\ref{aff27},\ref{aff16}}
\and M.~Brescia\orcid{0000-0001-9506-5680}\inst{\ref{aff28},\ref{aff29}}
\and J.~Brinchmann\orcid{0000-0003-4359-8797}\inst{\ref{aff30},\ref{aff31},\ref{aff32}}
\and S.~Camera\orcid{0000-0003-3399-3574}\inst{\ref{aff33},\ref{aff34},\ref{aff35}}
\and G.~Ca\~nas-Herrera\orcid{0000-0003-2796-2149}\inst{\ref{aff36}}
\and V.~Capobianco\orcid{0000-0002-3309-7692}\inst{\ref{aff35}}
\and C.~Carbone\orcid{0000-0003-0125-3563}\inst{\ref{aff37}}
\and V.~F.~Cardone\inst{\ref{aff38},\ref{aff39}}
\and J.~Carretero\orcid{0000-0002-3130-0204}\inst{\ref{aff40},\ref{aff41}}
\and F.~J.~Castander\orcid{0000-0001-7316-4573}\inst{\ref{aff42},\ref{aff43}}
\and M.~Castellano\orcid{0000-0001-9875-8263}\inst{\ref{aff38}}
\and G.~Castignani\orcid{0000-0001-6831-0687}\inst{\ref{aff17}}
\and S.~Cavuoti\orcid{0000-0002-3787-4196}\inst{\ref{aff29},\ref{aff44}}
\and K.~C.~Chambers\orcid{0000-0001-6965-7789}\inst{\ref{aff45}}
\and A.~Cimatti\inst{\ref{aff46}}
\and C.~Colodro-Conde\inst{\ref{aff47}}
\and G.~Congedo\orcid{0000-0003-2508-0046}\inst{\ref{aff48}}
\and C.~J.~Conselice\orcid{0000-0003-1949-7638}\inst{\ref{aff49}}
\and L.~Conversi\orcid{0000-0002-6710-8476}\inst{\ref{aff50},\ref{aff14}}
\and Y.~Copin\orcid{0000-0002-5317-7518}\inst{\ref{aff51}}
\and F.~Courbin\orcid{0000-0003-0758-6510}\inst{\ref{aff52},\ref{aff53},\ref{aff54}}
\and H.~M.~Courtois\orcid{0000-0003-0509-1776}\inst{\ref{aff55}}
\and J.-G.~Cuby\orcid{0000-0002-8767-1442}\inst{\ref{aff56},\ref{aff57}}
\and A.~Da~Silva\orcid{0000-0002-6385-1609}\inst{\ref{aff58},\ref{aff59}}
\and H.~Degaudenzi\orcid{0000-0002-5887-6799}\inst{\ref{aff60}}
\and G.~De~Lucia\orcid{0000-0002-6220-9104}\inst{\ref{aff12}}
\and H.~Dole\orcid{0000-0002-9767-3839}\inst{\ref{aff61}}
\and F.~Dubath\orcid{0000-0002-6533-2810}\inst{\ref{aff60}}
\and X.~Dupac\inst{\ref{aff14}}
\and S.~Escoffier\orcid{0000-0002-2847-7498}\inst{\ref{aff62}}
\and M.~Farina\orcid{0000-0002-3089-7846}\inst{\ref{aff63}}
\and R.~Farinelli\inst{\ref{aff17}}
\and S.~Farrens\orcid{0000-0002-9594-9387}\inst{\ref{aff8}}
\and F.~Faustini\orcid{0000-0001-6274-5145}\inst{\ref{aff38}}
\and S.~Ferriol\inst{\ref{aff51}}
\and F.~Finelli\orcid{0000-0002-6694-3269}\inst{\ref{aff17},\ref{aff64}}
\and S.~Fotopoulou\orcid{0000-0002-9686-254X}\inst{\ref{aff65}}
\and N.~Fourmanoit\orcid{0009-0005-6816-6925}\inst{\ref{aff62}}
\and M.~Frailis\orcid{0000-0002-7400-2135}\inst{\ref{aff12}}
\and E.~Franceschi\orcid{0000-0002-0585-6591}\inst{\ref{aff17}}
\and M.~Fumana\orcid{0000-0001-6787-5950}\inst{\ref{aff37}}
\and L.~Gabarra\orcid{0000-0002-8486-8856}\inst{\ref{aff5}}
\and S.~Galeotta\orcid{0000-0002-3748-5115}\inst{\ref{aff12}}
\and K.~George\orcid{0000-0002-1734-8455}\inst{\ref{aff66}}
\and W.~Gillard\orcid{0000-0003-4744-9748}\inst{\ref{aff62}}
\and B.~Gillis\orcid{0000-0002-4478-1270}\inst{\ref{aff48}}
\and C.~Giocoli\orcid{0000-0002-9590-7961}\inst{\ref{aff17},\ref{aff21}}
\and P.~G\'omez-Alvarez\orcid{0000-0002-8594-5358}\inst{\ref{aff67},\ref{aff14}}
\and J.~Gracia-Carpio\orcid{0000-0003-4689-3134}\inst{\ref{aff24}}
\and A.~Grazian\orcid{0000-0002-5688-0663}\inst{\ref{aff68}}
\and F.~Grupp\inst{\ref{aff24},\ref{aff25}}
\and L.~Guzzo\orcid{0000-0001-8264-5192}\inst{\ref{aff69},\ref{aff16},\ref{aff70}}
\and W.~G.~Hartley\orcid{0000-0002-4563-5101}\inst{\ref{aff60}}
\and S.~V.~H.~Haugan\orcid{0000-0001-9648-7260}\inst{\ref{aff71}}
\and S.~Hemmati\orcid{0000-0003-2226-5395}\inst{\ref{aff72}}
\and H.~Hoekstra\orcid{0000-0002-0641-3231}\inst{\ref{aff36}}
\and M.~S.~Holliman\inst{\ref{aff48}}
\and W.~Holmes\orcid{0009-0007-8554-4646}\inst{\ref{aff73}}
\and I.~M.~Hook\orcid{0000-0002-2960-978X}\inst{\ref{aff74}}
\and F.~Hormuth\inst{\ref{aff75}}
\and A.~Hornstrup\orcid{0000-0002-3363-0936}\inst{\ref{aff76},\ref{aff77}}
\and M.~Huertas-Company\orcid{0000-0002-1416-8483}\inst{\ref{aff47},\ref{aff78},\ref{aff79}}
\and K.~Jahnke\orcid{0000-0003-3804-2137}\inst{\ref{aff10}}
\and M.~Jhabvala\inst{\ref{aff80}}
\and S.~Kermiche\orcid{0000-0002-0302-5735}\inst{\ref{aff62}}
\and A.~Kiessling\orcid{0000-0002-2590-1273}\inst{\ref{aff73}}
\and R.~Kohley\inst{\ref{aff14}}
\and B.~Kubik\orcid{0009-0006-5823-4880}\inst{\ref{aff51}}
\and K.~Kuijken\orcid{0000-0002-3827-0175}\inst{\ref{aff36}}
\and M.~K\"ummel\orcid{0000-0003-2791-2117}\inst{\ref{aff25}}
\and M.~Kunz\orcid{0000-0002-3052-7394}\inst{\ref{aff81}}
\and H.~Kurki-Suonio\orcid{0000-0002-4618-3063}\inst{\ref{aff82},\ref{aff83}}
\and R.~Laureijs\inst{\ref{aff22}}
\and Q.~Le~Boulc'h\inst{\ref{aff84}}
\and A.~M.~C.~Le~Brun\orcid{0000-0002-0936-4594}\inst{\ref{aff85}}
\and D.~Le~Mignant\orcid{0000-0002-5339-5515}\inst{\ref{aff57}}
\and S.~Ligori\orcid{0000-0003-4172-4606}\inst{\ref{aff35}}
\and P.~B.~Lilje\orcid{0000-0003-4324-7794}\inst{\ref{aff71}}
\and V.~Lindholm\orcid{0000-0003-2317-5471}\inst{\ref{aff82},\ref{aff83}}
\and I.~Lloro\orcid{0000-0001-5966-1434}\inst{\ref{aff86}}
\and M.~Magliocchetti\orcid{0000-0001-9158-4838}\inst{\ref{aff63}}
\and G.~Mainetti\orcid{0000-0003-2384-2377}\inst{\ref{aff84}}
\and O.~Mansutti\orcid{0000-0001-5758-4658}\inst{\ref{aff12}}
\and S.~Marcin\inst{\ref{aff87}}
\and O.~Marggraf\orcid{0000-0001-7242-3852}\inst{\ref{aff6}}
\and M.~Martinelli\orcid{0000-0002-6943-7732}\inst{\ref{aff38},\ref{aff39}}
\and N.~Martinet\orcid{0000-0003-2786-7790}\inst{\ref{aff57}}
\and F.~Marulli\orcid{0000-0002-8850-0303}\inst{\ref{aff88},\ref{aff17},\ref{aff21}}
\and D.~C.~Masters\orcid{0000-0001-5382-6138}\inst{\ref{aff89}}
\and N.~Mauri\orcid{0000-0001-8196-1548}\inst{\ref{aff46},\ref{aff21}}
\and C.~J.~R.~McPartland\orcid{0000-0003-0639-025X}\inst{\ref{aff77},\ref{aff90}}
\and E.~Medinaceli\orcid{0000-0002-4040-7783}\inst{\ref{aff17}}
\and S.~Mei\orcid{0000-0002-2849-559X}\inst{\ref{aff91},\ref{aff92}}
\and M.~Melchior\inst{\ref{aff93}}
\and M.~Meneghetti\orcid{0000-0003-1225-7084}\inst{\ref{aff17},\ref{aff21}}
\and E.~Merlin\orcid{0000-0001-6870-8900}\inst{\ref{aff68}}
\and G.~Meylan\orcid{0000-0001-6503-0209}\inst{\ref{aff94}}
\and J.~J.~Mohr\orcid{0000-0002-6875-2087}\inst{\ref{aff66}}
\and P.~Monaco\orcid{0000-0003-2083-7564}\inst{\ref{aff95},\ref{aff12},\ref{aff18},\ref{aff13}}
\and A.~Mora\orcid{0000-0002-1922-8529}\inst{\ref{aff96}}
\and M.~Moresco\orcid{0000-0002-7616-7136}\inst{\ref{aff88},\ref{aff17}}
\and C.~Moretti\orcid{0000-0003-3314-8936}\inst{\ref{aff12},\ref{aff13},\ref{aff18}}
\and L.~Moscardini\orcid{0000-0002-3473-6716}\inst{\ref{aff88},\ref{aff17},\ref{aff21}}
\and E.~Munari\orcid{0000-0002-1751-5946}\inst{\ref{aff12},\ref{aff13}}
\and C.~Neissner\orcid{0000-0001-8524-4968}\inst{\ref{aff97},\ref{aff41}}
\and R.~C.~Nichol\orcid{0000-0003-0939-6518}\inst{\ref{aff15}}
\and S.-M.~Niemi\orcid{0009-0005-0247-0086}\inst{\ref{aff98}}
\and C.~Padilla\orcid{0000-0001-7951-0166}\inst{\ref{aff97}}
\and S.~Paltani\orcid{0000-0002-8108-9179}\inst{\ref{aff60}}
\and F.~Pasian\orcid{0000-0002-4869-3227}\inst{\ref{aff12}}
\and J.~A.~Peacock\orcid{0000-0002-1168-8299}\inst{\ref{aff48}}
\and W.~J.~Percival\orcid{0000-0002-0644-5727}\inst{\ref{aff99},\ref{aff100},\ref{aff101}}
\and V.~Pettorino\orcid{0000-0002-4203-9320}\inst{\ref{aff98}}
\and A.~Pezzotta\orcid{0000-0003-0726-2268}\inst{\ref{aff16}}
\and S.~Pires\orcid{0000-0002-0249-2104}\inst{\ref{aff8}}
\and G.~Polenta\orcid{0000-0003-4067-9196}\inst{\ref{aff102}}
\and M.~Poncet\inst{\ref{aff9}}
\and L.~A.~Popa\inst{\ref{aff103}}
\and L.~Pozzetti\orcid{0000-0001-7085-0412}\inst{\ref{aff17}}
\and F.~Raison\orcid{0000-0002-7819-6918}\inst{\ref{aff24}}
\and R.~Rebolo\orcid{0000-0003-3767-7085}\inst{\ref{aff47},\ref{aff104},\ref{aff105}}
\and A.~Refregier\inst{\ref{aff106}}
\and A.~Renzi\orcid{0000-0001-9856-1970}\inst{\ref{aff107},\ref{aff108},\ref{aff17}}
\and J.~Rhodes\orcid{0000-0002-4485-8549}\inst{\ref{aff73}}
\and G.~Riccio\inst{\ref{aff29}}
\and I.~Risso\orcid{0000-0003-2525-7761}\inst{\ref{aff26},\ref{aff27},\ref{aff16}}
\and E.~Romelli\orcid{0000-0003-3069-9222}\inst{\ref{aff12}}
\and M.~Roncarelli\orcid{0000-0001-9587-7822}\inst{\ref{aff17}}
\and C.~Rosset\orcid{0000-0003-0286-2192}\inst{\ref{aff91}}
\and B.~Rusholme\orcid{0000-0001-7648-4142}\inst{\ref{aff72}}
\and R.~Saglia\orcid{0000-0003-0378-7032}\inst{\ref{aff25},\ref{aff24}}
\and Z.~Sakr\orcid{0000-0002-4823-3757}\inst{\ref{aff109},\ref{aff110},\ref{aff111}}
\and D.~Sapone\orcid{0000-0001-7089-4503}\inst{\ref{aff112}}
\and B.~Sartoris\orcid{0000-0003-1337-5269}\inst{\ref{aff24},\ref{aff12}}
\and M.~Sauvage\orcid{0000-0002-0809-2574}\inst{\ref{aff8}}
\and P.~Schneider\orcid{0000-0001-8561-2679}\inst{\ref{aff6}}
\and T.~Schrabback\orcid{0000-0002-6987-7834}\inst{\ref{aff113}}
\and M.~Scodeggio\inst{\ref{aff37}}
\and A.~Secroun\orcid{0000-0003-0505-3710}\inst{\ref{aff62}}
\and M.~Seiffert\orcid{0000-0002-7536-9393}\inst{\ref{aff73}}
\and E.~Sihvola\orcid{0000-0003-1804-7715}\inst{\ref{aff114}}
\and C.~Sirignano\orcid{0000-0002-0995-7146}\inst{\ref{aff107},\ref{aff108}}
\and G.~Sirri\orcid{0000-0003-2626-2853}\inst{\ref{aff21}}
\and L.~Stanco\orcid{0000-0002-9706-5104}\inst{\ref{aff108}}
\and P.~Tallada-Cresp\'{i}\orcid{0000-0002-1336-8328}\inst{\ref{aff40},\ref{aff41}}
\and A.~N.~Taylor\inst{\ref{aff48}}
\and H.~I.~Teplitz\orcid{0000-0002-7064-5424}\inst{\ref{aff89}}
\and I.~Tereno\orcid{0000-0002-4537-6218}\inst{\ref{aff58},\ref{aff115}}
\and N.~Tessore\orcid{0000-0002-9696-7931}\inst{\ref{aff2}}
\and S.~Toft\orcid{0000-0003-3631-7176}\inst{\ref{aff116},\ref{aff90}}
\and R.~Toledo-Moreo\orcid{0000-0002-2997-4859}\inst{\ref{aff117},\ref{aff118}}
\and F.~Torradeflot\orcid{0000-0003-1160-1517}\inst{\ref{aff41},\ref{aff40}}
\and A.~Tsyganov\inst{\ref{aff119}}
\and I.~Tutusaus\orcid{0000-0002-3199-0399}\inst{\ref{aff42},\ref{aff43},\ref{aff110}}
\and E.~A.~Valentijn\inst{\ref{aff22}}
\and J.~Valiviita\orcid{0000-0001-6225-3693}\inst{\ref{aff82},\ref{aff83}}
\and T.~Vassallo\orcid{0000-0001-6512-6358}\inst{\ref{aff12},\ref{aff66}}
\and A.~Veropalumbo\orcid{0000-0003-2387-1194}\inst{\ref{aff16},\ref{aff27},\ref{aff26}}
\and Y.~Wang\orcid{0000-0002-4749-2984}\inst{\ref{aff72}}
\and J.~Weller\orcid{0000-0002-8282-2010}\inst{\ref{aff25},\ref{aff24}}
\and G.~Zamorani\orcid{0000-0002-2318-301X}\inst{\ref{aff17}}
\and F.~M.~Zerbi\orcid{0000-0002-9996-973X}\inst{\ref{aff16}}
\and E.~Zucca\orcid{0000-0002-5845-8132}\inst{\ref{aff17}}
\and V.~Allevato\orcid{0000-0001-7232-5152}\inst{\ref{aff29}}
\and M.~Ballardini\orcid{0000-0003-4481-3559}\inst{\ref{aff120},\ref{aff121},\ref{aff17}}
\and C.~Benoist\inst{\ref{aff122}}
\and P.~Bergamini\orcid{0000-0003-1383-9414}\inst{\ref{aff17}}
\and S.~Borgani\orcid{0000-0001-6151-6439}\inst{\ref{aff95},\ref{aff13},\ref{aff12},\ref{aff18},\ref{aff123}}
\and E.~Bozzo\orcid{0000-0002-8201-1525}\inst{\ref{aff60}}
\and C.~Burigana\orcid{0000-0002-3005-5796}\inst{\ref{aff124},\ref{aff64}}
\and R.~Cabanac\orcid{0000-0001-6679-2600}\inst{\ref{aff110}}
\and M.~Calabrese\orcid{0000-0002-2637-2422}\inst{\ref{aff125},\ref{aff37}}
\and A.~Cappi\inst{\ref{aff122},\ref{aff17}}
\and F.~Caro\orcid{0009-0003-1053-0507}\inst{\ref{aff38}}
\and T.~Castro\orcid{0000-0002-6292-3228}\inst{\ref{aff126},\ref{aff12},\ref{aff13}}
\and D.~Di~Ferdinando\inst{\ref{aff21}}
\and J.~A.~Escartin~Vigo\inst{\ref{aff24}}
\and G.~Fabbian\orcid{0000-0002-3255-4695}\inst{\ref{aff61}}
\and J.~Garc\'ia-Bellido\orcid{0000-0002-9370-8360}\inst{\ref{aff109}}
\and T.~Gasparetto\orcid{0000-0002-7913-4866}\inst{\ref{aff38}}
\and V.~Gautard\inst{\ref{aff127}}
\and A.~Gregorio\orcid{0000-0003-4028-8785}\inst{\ref{aff95},\ref{aff12},\ref{aff18}}
\and C.~M.~Gutierrez\orcid{0000-0001-7854-783X}\inst{\ref{aff47},\ref{aff105}}
\and A.~Hall\orcid{0000-0002-3139-8651}\inst{\ref{aff48}}
\and A.~Kashlinsky\orcid{0000-0003-2156-078X}\inst{\ref{aff128},\ref{aff129}}
\and A.~Loureiro\orcid{0000-0002-4371-0876}\inst{\ref{aff130},\ref{aff131}}
\and J.~Macias-Perez\orcid{0000-0002-5385-2763}\inst{\ref{aff132}}
\and C.~Mancini\orcid{0000-0002-4297-0561}\inst{\ref{aff37}}
\and R.~Maoli\orcid{0000-0002-6065-3025}\inst{\ref{aff133},\ref{aff38}}
\and R.~B.~Metcalf\orcid{0000-0003-3167-2574}\inst{\ref{aff88},\ref{aff17}}
\and M.~P\"ontinen\orcid{0000-0001-5442-2530}\inst{\ref{aff82}}
\and E.~Sarpa\orcid{0000-0002-1256-655X}\inst{\ref{aff12}}
\and V.~Scottez\orcid{0009-0008-3864-940X}\inst{\ref{aff134},\ref{aff135}}
\and M.~Sereno\orcid{0000-0003-0302-0325}\inst{\ref{aff17},\ref{aff21}}
\and K.~Tanidis\orcid{0000-0001-9843-5130}\inst{\ref{aff136}}
\and M.~Tenti\orcid{0000-0002-4254-5901}\inst{\ref{aff21}}
\and M.~Tucci\inst{\ref{aff60}}
\and M.~Viel\orcid{0000-0002-2642-5707}\inst{\ref{aff13},\ref{aff12},\ref{aff19},\ref{aff18},\ref{aff123}}
\and J.~R.~Weaver\orcid{0000-0003-1614-196X}\inst{\ref{aff137}}
\and M.~Wiesmann\orcid{0009-0000-8199-5860}\inst{\ref{aff71}}
\and J.~A.~Acevedo~Barroso\orcid{0000-0002-9654-1711}\inst{\ref{aff73}}
\and Y.~Akrami\orcid{0000-0002-2407-7956}\inst{\ref{aff109},\ref{aff138}}
\and I.~T.~Andika\orcid{0000-0001-6102-9526}\inst{\ref{aff25}}
\and G.~Angora\orcid{0000-0002-0316-6562}\inst{\ref{aff29},\ref{aff120}}
\and S.~Anselmi\orcid{0000-0002-3579-9583}\inst{\ref{aff108},\ref{aff107},\ref{aff139}}
\and M.~Archidiacono\orcid{0000-0003-4952-9012}\inst{\ref{aff69},\ref{aff70}}
\and G.~Aric\`o\orcid{0000-0002-2802-2928}\inst{\ref{aff21}}
\and F.~Atrio-Barandela\orcid{0000-0002-2130-2513}\inst{\ref{aff140}}
\and M.~Aubert\orcid{0009-0002-7667-8814}\inst{\ref{aff51}}
\and M.~Baes\orcid{0000-0002-3930-2757}\inst{\ref{aff141}}
\and L.~Bazzanini\orcid{0000-0003-0727-0137}\inst{\ref{aff120},\ref{aff17}}
\and D.~Bertacca\orcid{0000-0002-2490-7139}\inst{\ref{aff107},\ref{aff68},\ref{aff108}}
\and M.~Bethermin\orcid{0000-0002-3915-2015}\inst{\ref{aff142}}
\and F.~Beutler\orcid{0000-0003-0467-5438}\inst{\ref{aff48}}
\and L.~Bisigello\orcid{0000-0003-0492-4924}\inst{\ref{aff68}}
\and A.~Blanchard\orcid{0000-0001-8555-9003}\inst{\ref{aff110}}
\and L.~Blot\orcid{0000-0002-9622-7167}\inst{\ref{aff143},\ref{aff85}}
\and F.~Bouch\`e\orcid{0000-0002-4663-1786}\inst{\ref{aff144},\ref{aff44}}
\and M.~L.~Brown\orcid{0000-0002-0370-8077}\inst{\ref{aff49}}
\and S.~Bruton\orcid{0000-0002-6503-5218}\inst{\ref{aff145}}
\and F.~Buitrago\orcid{0000-0002-2861-9812}\inst{\ref{aff146},\ref{aff147},\ref{aff115}}
\and A.~Calabro\orcid{0000-0003-2536-1614}\inst{\ref{aff38}}
\and B.~Camacho~Quevedo\orcid{0000-0002-8789-4232}\inst{\ref{aff13},\ref{aff19},\ref{aff12}}
\and P.~Carrilho\orcid{0000-0003-1339-0194}\inst{\ref{aff148}}
\and C.~S.~Carvalho\inst{\ref{aff115}}
\and F.~Cogato\orcid{0000-0003-4632-6113}\inst{\ref{aff17}}
\and T.~E.~Collett\orcid{0000-0001-5564-3140}\inst{\ref{aff149}}
\and S.~Conseil\orcid{0000-0002-3657-4191}\inst{\ref{aff150},\ref{aff51}}
\and A.~R.~Cooray\orcid{0000-0002-3892-0190}\inst{\ref{aff151}}
\and P.~Corcho-Caballero\orcid{0000-0001-6327-7080}\inst{\ref{aff22}}
\and M.~Correnti\orcid{0000-0001-6464-3257}\inst{\ref{aff38},\ref{aff102}}
\and B.~Csizi\orcid{0000-0003-3227-6581}\inst{\ref{aff113}}
\and O.~Cucciati\orcid{0000-0002-9336-7551}\inst{\ref{aff17}}
\and H.~Dannerbauer\orcid{0000-0001-7147-3575}\inst{\ref{aff47},\ref{aff105}}
\and T.~de~Boer\orcid{0000-0001-5486-2747}\inst{\ref{aff45}}
\and G.~Desprez\orcid{0000-0001-8325-1742}\inst{\ref{aff22}}
\and A.~D\'iaz-S\'anchez\orcid{0000-0003-0748-4768}\inst{\ref{aff152}}
\and S.~Di~Domizio\orcid{0000-0003-2863-5895}\inst{\ref{aff26},\ref{aff27}}
\and J.~M.~Diego\orcid{0000-0001-9065-3926}\inst{\ref{aff153}}
\and P.~Dimauro\orcid{0000-0001-7399-2854}\inst{\ref{aff154},\ref{aff38}}
\and P.-A.~Duc\orcid{0000-0003-3343-6284}\inst{\ref{aff142}}
\and V.~Duret\orcid{0009-0009-0383-4960}\inst{\ref{aff62}}
\and A.~Enia\orcid{0000-0002-0200-2857}\inst{\ref{aff127},\ref{aff17}}
\and Y.~Fang\orcid{0000-0002-0334-6950}\inst{\ref{aff25}}
\and A.~Farina\orcid{0009-0000-3420-929X}\inst{\ref{aff16},\ref{aff27}}
\and A.~M.~N.~Ferguson\orcid{0000-0001-7934-1278}\inst{\ref{aff48}}
\and A.~Finoguenov\orcid{0000-0002-4606-5403}\inst{\ref{aff82}}
\and A.~Franco\orcid{0000-0002-4761-366X}\inst{\ref{aff155},\ref{aff156},\ref{aff157}}
\and A.~Fumagalli\orcid{0009-0004-0300-2535}\inst{\ref{aff12}}
\and Y.~Fu\orcid{0000-0002-0759-0504}\inst{\ref{aff36},\ref{aff22}}
\and K.~Ganga\orcid{0000-0001-8159-8208}\inst{\ref{aff91}}
\and R.~Gavazzi\orcid{0000-0002-5540-6935}\inst{\ref{aff57},\ref{aff1}}
\and E.~Gaztanaga\orcid{0000-0001-9632-0815}\inst{\ref{aff42},\ref{aff43},\ref{aff149}}
\and Z.~Ghaffari\orcid{0000-0002-6467-8078}\inst{\ref{aff12},\ref{aff13}}
\and F.~Giacomini\orcid{0000-0002-3129-2814}\inst{\ref{aff21}}
\and F.~Gianotti\orcid{0000-0003-4666-119X}\inst{\ref{aff17}}
\and G.~Gozaliasl\orcid{0000-0002-0236-919X}\inst{\ref{aff158},\ref{aff82}}
\and A.~Gruppuso\orcid{0000-0001-9272-5292}\inst{\ref{aff17},\ref{aff21}}
\and M.~Guidi\orcid{0000-0001-9408-1101}\inst{\ref{aff20},\ref{aff17}}
\and N.~A.~Hatch\orcid{0000-0001-5600-0534}\inst{\ref{aff159}}
\and C.~Hern\'andez-Monteagudo\orcid{0000-0001-5471-9166}\inst{\ref{aff105},\ref{aff47}}
\and T.~Heuberger\orcid{0009-0009-3474-8467}\inst{\ref{aff87}}
\and H.~Hildebrandt\orcid{0000-0002-9814-3338}\inst{\ref{aff160}}
\and J.~Hjorth\orcid{0000-0002-4571-2306}\inst{\ref{aff161}}
\and L.~K.~Hunt\orcid{0000-0001-9162-2371}\inst{\ref{aff162}}
\and H.~Jansen\orcid{0009-0002-1332-7742}\inst{\ref{aff113}}
\and M.~Jauzac\orcid{0000-0003-1974-8732}\inst{\ref{aff3},\ref{aff4},\ref{aff163},\ref{aff164},\ref{aff110}}
\and S.~Joudaki\orcid{0000-0001-8820-673X}\inst{\ref{aff40},\ref{aff149}}
\and J.~J.~E.~Kajava\orcid{0000-0002-3010-8333}\inst{\ref{aff165},\ref{aff166},\ref{aff167}}
\and Y.~Kang\orcid{0009-0000-8588-7250}\inst{\ref{aff113}}
\and V.~Kansal\orcid{0000-0002-4008-6078}\inst{\ref{aff168},\ref{aff169}}
\and D.~Karagiannis\orcid{0000-0002-4927-0816}\inst{\ref{aff120},\ref{aff170}}
\and J.~Kim\orcid{0000-0003-2776-2761}\inst{\ref{aff5}}
\and C.~C.~Kirkpatrick\inst{\ref{aff114}}
\and A.~Kov\'acs\orcid{0000-0002-5825-579X}\inst{\ref{aff171},\ref{aff172}}
\and I.~Kova{\v{c}}i{\'{c}}\orcid{0000-0001-6751-3263}\inst{\ref{aff141}}
\and K.~Koyama\orcid{0000-0001-6727-6915}\inst{\ref{aff149}}
\and S.~Kruk\orcid{0000-0001-8010-8879}\inst{\ref{aff14}}
\and M.~C.~Lam\orcid{0000-0002-9347-2298}\inst{\ref{aff48}}
\and M.~Lattanzi\orcid{0000-0003-1059-2532}\inst{\ref{aff121}}
\and S.~Lee\orcid{0000-0002-8289-740X}\inst{\ref{aff173},\ref{aff73}}
\and J.~Le~Graet\orcid{0000-0001-6523-7971}\inst{\ref{aff62}}
\and L.~Legrand\orcid{0000-0003-0610-5252}\inst{\ref{aff174},\ref{aff61}}
\and M.~Lembo\orcid{0000-0002-5271-5070}\inst{\ref{aff1}}
\and J.~Lesgourgues\orcid{0000-0001-7627-353X}\inst{\ref{aff175}}
\and T.~I.~Liaudat\orcid{0000-0002-9104-314X}\inst{\ref{aff176}}
\and L.~Linke\orcid{0000-0002-2622-8113}\inst{\ref{aff113}}
\and S.~J.~Liu\orcid{0000-0001-7680-2139}\inst{\ref{aff63}}
\and G.~Maggio\orcid{0000-0003-4020-4836}\inst{\ref{aff12}}
\and F.~Mannucci\orcid{0000-0002-4803-2381}\inst{\ref{aff162}}
\and F.~R.~Marleau\orcid{0000-0002-1442-2947}\inst{\ref{aff113}}
\and C.~J.~A.~P.~Martins\orcid{0000-0002-4886-9261}\inst{\ref{aff177},\ref{aff30}}
\and M.~Migliaccio\inst{\ref{aff178},\ref{aff179}}
\and M.~Miluzio\inst{\ref{aff14},\ref{aff180}}
\and G.~Morgante\inst{\ref{aff17}}
\and S.~Nadathur\orcid{0000-0001-9070-3102}\inst{\ref{aff149}}
\and K.~Naidoo\orcid{0000-0002-9182-1802}\inst{\ref{aff149},\ref{aff10}}
\and A.~Navarro-Alsina\orcid{0000-0002-3173-2592}\inst{\ref{aff6}}
\and A.~Nersesian\orcid{0000-0001-6843-409X}\inst{\ref{aff181},\ref{aff141}}
\and S.~Nesseris\orcid{0000-0002-0567-0324}\inst{\ref{aff109}}
\and L.~Nicastro\orcid{0000-0001-8534-6788}\inst{\ref{aff17}}
\and F.~Oppizzi\orcid{0000-0003-3904-8370}\inst{\ref{aff27},\ref{aff108},\ref{aff107}}
\and F.~Pace\orcid{0000-0001-8039-0480}\inst{\ref{aff33},\ref{aff34},\ref{aff35},\ref{aff59}}
\and L.~Pagano\orcid{0000-0003-1820-5998}\inst{\ref{aff120},\ref{aff121}}
\and D.~Paoletti\orcid{0000-0003-4761-6147}\inst{\ref{aff17},\ref{aff64}}
\and F.~Passalacqua\orcid{0000-0002-8606-4093}\inst{\ref{aff108}}
\and L.~Patrizii\inst{\ref{aff21}}
\and C.~Pattison\orcid{0000-0003-3272-2617}\inst{\ref{aff149}}
\and A.~Pisani\orcid{0000-0002-6146-4437}\inst{\ref{aff62}}
\and L.~Pizzuti\orcid{0000-0001-5654-7580}\inst{\ref{aff182}}
\and J.~J.~Popp\orcid{0000-0002-3724-1727}\inst{\ref{aff183}}
\and D.~Potter\orcid{0000-0002-0757-5195}\inst{\ref{aff184}}
\and M.~Poulain\orcid{0000-0002-7664-4510}\inst{\ref{aff185}}
\and G.~W.~Pratt\inst{\ref{aff8}}
\and S.~Quai\orcid{0000-0002-0449-8163}\inst{\ref{aff88},\ref{aff17}}
\and M.~Radovich\orcid{0000-0002-3585-866X}\inst{\ref{aff68}}
\and F.~Ricci\orcid{0000-0001-5742-5980}\inst{\ref{aff186},\ref{aff38}}
\and G.~Rodighiero\orcid{0000-0002-9415-2296}\inst{\ref{aff107},\ref{aff68}}
\and W.~Roster\orcid{0000-0002-9149-6528}\inst{\ref{aff24}}
\and M.~Sahl\'en\orcid{0000-0003-0973-4804}\inst{\ref{aff187}}
\and D.~B.~Sanders\orcid{0000-0002-1233-9998}\inst{\ref{aff45}}
\and A.~Schneider\orcid{0000-0001-7055-8104}\inst{\ref{aff184}}
\and D.~Sciotti\orcid{0009-0008-4519-2620}\inst{\ref{aff38},\ref{aff39}}
\and E.~Sellentin\orcid{0009-0002-2655-3458}\inst{\ref{aff188},\ref{aff36}}
\and S.~Serjeant\orcid{0000-0002-0517-7943}\inst{\ref{aff183}}
\and F.~Shankar\orcid{0000-0001-8973-5051}\inst{\ref{aff189}}
\and I.~Szapudi\orcid{0000-0003-2274-0301}\inst{\ref{aff45}}
\and M.~Talia\orcid{0000-0003-4352-2063}\inst{\ref{aff88},\ref{aff17}}
\and F.~Tarsitano\orcid{0000-0002-5919-0238}\inst{\ref{aff190},\ref{aff106},\ref{aff60}}
\and G.~Testera\orcid{0000-0003-2970-766X}\inst{\ref{aff27}}
\and M.~Tewes\orcid{0000-0002-1155-8689}\inst{\ref{aff6}}
\and R.~Teyssier\orcid{0000-0001-7689-0933}\inst{\ref{aff191}}
\and S.~Tosi\orcid{0000-0002-7275-9193}\inst{\ref{aff26},\ref{aff16},\ref{aff27}}
\and A.~Troja\orcid{0000-0003-0239-4595}\inst{\ref{aff12}}
\and C.~Uhlemann\orcid{0000-0001-7831-1579}\inst{\ref{aff192},\ref{aff193}}
\and M.~Urbano\orcid{0000-0001-5640-0650}\inst{\ref{aff159},\ref{aff142}}
\and C.~Valieri\inst{\ref{aff21}}
\and A.~Venhola\orcid{0000-0001-6071-4564}\inst{\ref{aff185}}
\and D.~Vergani\orcid{0000-0003-0898-2216}\inst{\ref{aff17}}
\and P.~Vielzeuf\orcid{0000-0003-2035-9339}\inst{\ref{aff62}}
\and A.~Viitanen\orcid{0000-0001-9383-786X}\inst{\ref{aff60},\ref{aff114},\ref{aff38}}
\and S.~Vinciguerra\orcid{0009-0005-4018-3184}\inst{\ref{aff57}}
\and N.~A.~Walton\orcid{0000-0003-3983-8778}\inst{\ref{aff11}}
\and L.~Wang\orcid{0000-0002-6736-9158}\inst{\ref{aff194},\ref{aff22}}
\and A.~H.~Wright\orcid{0000-0001-7363-7932}\inst{\ref{aff160}}
\and L.~Zalesky\orcid{0000-0001-5680-2326}\inst{\ref{aff45}}
\and D.~Scott\orcid{0000-0002-6878-9840}\inst{\ref{aff195}}}
										   
\institute{Institut d'Astrophysique de Paris, UMR 7095, CNRS, and Sorbonne Universit\'e, 98 bis boulevard Arago, 75014 Paris, France\label{aff1}
\and
Mullard Space Science Laboratory, University College London, Holmbury St Mary, Dorking, Surrey RH5 6NT, UK\label{aff2}
\and
Department of Physics, Centre for Extragalactic Astronomy, Durham University, South Road, Durham, DH1 3LE, UK\label{aff3}
\and
Department of Physics, Institute for Computational Cosmology, Durham University, South Road, Durham, DH1 3LE, UK\label{aff4}
\and
Department of Physics, University of Oxford, Keble Road, Oxford OX1 3RH, UK\label{aff5}
\and
Universit\"at Bonn, Argelander-Institut f\"ur Astronomie, Auf dem H\"ugel 71, 53121 Bonn, Germany\label{aff6}
\and
Centre for Electronic Imaging, Open University, Walton Hall, Milton Keynes, MK7~6AA, UK\label{aff7}
\and
Universit\'e Paris-Saclay, Universit\'e Paris Cit\'e, CEA, CNRS, AIM, 91191, Gif-sur-Yvette, France\label{aff8}
\and
Centre National d'Etudes Spatiales -- Centre spatial de Toulouse, 18 avenue Edouard Belin, 31401 Toulouse Cedex 9, France\label{aff9}
\and
Max-Planck-Institut f\"ur Astronomie, K\"onigstuhl 17, 69117 Heidelberg, Germany\label{aff10}
\and
Institute of Astronomy, University of Cambridge, Madingley Road, Cambridge CB3 0HA, UK\label{aff11}
\and
INAF-Osservatorio Astronomico di Trieste, Via G. B. Tiepolo 11, 34143 Trieste, Italy\label{aff12}
\and
IFPU, Institute for Fundamental Physics of the Universe, via Beirut 2, 34151 Trieste, Italy\label{aff13}
\and
ESAC/ESA, Camino Bajo del Castillo, s/n., Urb. Villafranca del Castillo, 28692 Villanueva de la Ca\~nada, Madrid, Spain\label{aff14}
\and
School of Mathematics and Physics, University of Surrey, Guildford, Surrey, GU2 7XH, UK\label{aff15}
\and
INAF-Osservatorio Astronomico di Brera, Via Brera 28, 20122 Milano, Italy\label{aff16}
\and
INAF-Osservatorio di Astrofisica e Scienza dello Spazio di Bologna, Via Piero Gobetti 93/3, 40129 Bologna, Italy\label{aff17}
\and
INFN, Sezione di Trieste, Via Valerio 2, 34127 Trieste TS, Italy\label{aff18}
\and
SISSA, International School for Advanced Studies, Via Bonomea 265, 34136 Trieste TS, Italy\label{aff19}
\and
Dipartimento di Fisica e Astronomia, Universit\`a di Bologna, Via Gobetti 93/2, 40129 Bologna, Italy\label{aff20}
\and
INFN-Sezione di Bologna, Viale Berti Pichat 6/2, 40127 Bologna, Italy\label{aff21}
\and
Kapteyn Astronomical Institute, University of Groningen, PO Box 800, 9700 AV Groningen, The Netherlands\label{aff22}
\and
ATG Europe BV, Huygensstraat 34, 2201 DK Noordwijk, The Netherlands\label{aff23}
\and
Max Planck Institute for Extraterrestrial Physics, Giessenbachstr. 1, 85748 Garching, Germany\label{aff24}
\and
Universit\"ats-Sternwarte M\"unchen, Fakult\"at f\"ur Physik, Ludwig-Maximilians-Universit\"at M\"unchen, Scheinerstr.~1, 81679 M\"unchen, Germany\label{aff25}
\and
Dipartimento di Fisica, Universit\`a di Genova, Via Dodecaneso 33, 16146, Genova, Italy\label{aff26}
\and
INFN-Sezione di Genova, Via Dodecaneso 33, 16146, Genova, Italy\label{aff27}
\and
Department of Physics "E. Pancini", University Federico II, Via Cinthia 6, 80126, Napoli, Italy\label{aff28}
\and
INAF-Osservatorio Astronomico di Capodimonte, Via Moiariello 16, 80131 Napoli, Italy\label{aff29}
\and
Instituto de Astrof\'isica e Ci\^encias do Espa\c{c}o, Universidade do Porto, CAUP, Rua das Estrelas, PT4150-762 Porto, Portugal\label{aff30}
\and
Faculdade de Ci\^encias da Universidade do Porto, Rua do Campo de Alegre, 4150-007 Porto, Portugal\label{aff31}
\and
European Southern Observatory, Karl-Schwarzschild-Str.~2, 85748 Garching, Germany\label{aff32}
\and
Dipartimento di Fisica, Universit\`a degli Studi di Torino, Via P. Giuria 1, 10125 Torino, Italy\label{aff33}
\and
INFN-Sezione di Torino, Via P. Giuria 1, 10125 Torino, Italy\label{aff34}
\and
INAF-Osservatorio Astrofisico di Torino, Via Osservatorio 20, 10025 Pino Torinese (TO), Italy\label{aff35}
\and
Leiden Observatory, Leiden University, Einsteinweg 55, 2333 CC Leiden, The Netherlands\label{aff36}
\and
INAF-IASF Milano, Via Alfonso Corti 12, 20133 Milano, Italy\label{aff37}
\and
INAF-Osservatorio Astronomico di Roma, Via Frascati 33, 00078 Monteporzio Catone, Italy\label{aff38}
\and
INFN-Sezione di Roma, Piazzale Aldo Moro, 2 - c/o Dipartimento di Fisica, Edificio G. Marconi, 00185 Roma, Italy\label{aff39}
\and
Centro de Investigaciones Energ\'eticas, Medioambientales y Tecnol\'ogicas (CIEMAT), Avenida Complutense 40, 28040 Madrid, Spain\label{aff40}
\and
Port d'Informaci\'{o} Cient\'{i}fica, Campus UAB, C. Albareda s/n, 08193 Bellaterra (Barcelona), Spain\label{aff41}
\and
Institute of Space Sciences (ICE, CSIC), Campus UAB, Carrer de Can Magrans, s/n, 08193 Barcelona, Spain\label{aff42}
\and
Institut d'Estudis Espacials de Catalunya (IEEC),  Edifici RDIT, Campus UPC, 08860 Castelldefels, Barcelona, Spain\label{aff43}
\and
INFN -- Sezione di Napoli, Via Cinthia 6, 80126, Napoli, Italy\label{aff44}
\and
Institute for Astronomy, University of Hawaii, 2680 Woodlawn Drive, Honolulu, HI 96822, USA\label{aff45}
\and
Dipartimento di Fisica e Astronomia "Augusto Righi" - Alma Mater Studiorum Universit\`a di Bologna, Viale Berti Pichat 6/2, 40127 Bologna, Italy\label{aff46}
\and
Instituto de Astrof\'{\i}sica de Canarias, E-38205 La Laguna, Tenerife, Spain\label{aff47}
\and
Institute for Astronomy, University of Edinburgh, Royal Observatory, Blackford Hill, Edinburgh EH9 3HJ, UK\label{aff48}
\and
Jodrell Bank Centre for Astrophysics, Department of Physics and Astronomy, University of Manchester, Oxford Road, Manchester M13 9PL, UK\label{aff49}
\and
European Space Agency/ESRIN, Largo Galileo Galilei 1, 00044 Frascati, Roma, Italy\label{aff50}
\and
Universit\'e Lyon 1, CNRS/IN2P3, IP2I Lyon, Villeurbanne, France\label{aff51}
\and
Institut de Ci\`{e}ncies del Cosmos (ICCUB), Universitat de Barcelona (IEEC-UB), Mart\'{i} i Franqu\`{e}s 1, 08028 Barcelona, Spain\label{aff52}
\and
Instituci\'o Catalana de Recerca i Estudis Avan\c{c}ats (ICREA), Passeig de Llu\'{\i}s Companys 23, 08010 Barcelona, Spain\label{aff53}
\and
Institut de Ciencies de l'Espai (IEEC-CSIC), Campus UAB, Carrer de Can Magrans, s/n Cerdanyola del Vall\'es, 08193 Barcelona, Spain\label{aff54}
\and
Universit\'e Lyon 1, CNRS/IN2P3, IUF, IP2I Lyon, Villeurbanne, France \label{aff55}
\and
Canada-France-Hawaii Telescope, 65-1238 Mamalahoa Hwy, Kamuela, HI 96743, USA\label{aff56}
\and
Aix-Marseille Universit\'e, CNRS, CNES, LAM, Marseille, France\label{aff57}
\and
Departamento de F\'isica, Faculdade de Ci\^encias, Universidade de Lisboa, Edif\'icio C8, Campo Grande, PT1749-016 Lisboa, Portugal\label{aff58}
\and
Instituto de Astrof\'isica e Ci\^encias do Espa\c{c}o, Faculdade de Ci\^encias, Universidade de Lisboa, Campo Grande, 1749-016 Lisboa, Portugal\label{aff59}
\and
Department of Astronomy, University of Geneva, ch. d'Ecogia 16, 1290 Versoix, Switzerland\label{aff60}
\and
Universit\'e Paris-Saclay, CNRS, Institut d'astrophysique spatiale, 91405, Orsay, France\label{aff61}
\and
Aix-Marseille Universit\'e, CNRS/IN2P3, CPPM, Marseille, France\label{aff62}
\and
INAF-Istituto di Astrofisica e Planetologia Spaziali, via del Fosso del Cavaliere, 100, 00100 Roma, Italy\label{aff63}
\and
INFN-Bologna, Via Irnerio 46, 40126 Bologna, Italy\label{aff64}
\and
School of Physics, HH Wills Physics Laboratory, University of Bristol, Tyndall Avenue, Bristol, BS8 1TL, UK\label{aff65}
\and
University Observatory, LMU Faculty of Physics, Scheinerstr.~1, 81679 Munich, Germany\label{aff66}
\and
FRACTAL S.L.N.E., calle Tulip\'an 2, Portal 13 1A, 28231, Las Rozas de Madrid, Spain\label{aff67}
\and
INAF-Osservatorio Astronomico di Padova, Via dell'Osservatorio 5, 35122 Padova, Italy\label{aff68}
\and
Dipartimento di Fisica "Aldo Pontremoli", Universit\`a degli Studi di Milano, Via Celoria 16, 20133 Milano, Italy\label{aff69}
\and
INFN-Sezione di Milano, Via Celoria 16, 20133 Milano, Italy\label{aff70}
\and
Institute of Theoretical Astrophysics, University of Oslo, P.O. Box 1029 Blindern, 0315 Oslo, Norway\label{aff71}
\and
Caltech/IPAC, 1200 E. California Blvd., Pasadena, CA 91125, USA\label{aff72}
\and
Jet Propulsion Laboratory, California Institute of Technology, 4800 Oak Grove Drive, Pasadena, CA, 91109, USA\label{aff73}
\and
School of Physics and Astronomy, Lancaster University, Lancaster, LA1 4YB, UK\label{aff74}
\and
Felix Hormuth Engineering, Goethestr. 17, 69181 Leimen, Germany\label{aff75}
\and
Technical University of Denmark, Elektrovej 327, 2800 Kgs. Lyngby, Denmark\label{aff76}
\and
Cosmic Dawn Center (DAWN), Denmark\label{aff77}
\and
Universit\'e PSL, Observatoire de Paris, Sorbonne Universit\'e, CNRS, LERMA, 75014, Paris, France\label{aff78}
\and
Universit\'e Paris-Cit\'e, 5 Rue Thomas Mann, 75013, Paris, France\label{aff79}
\and
NASA Goddard Space Flight Center, Greenbelt, MD 20771, USA\label{aff80}
\and
Universit\'e de Gen\`eve, D\'epartement de Physique Th\'eorique and Centre for Astroparticle Physics, 24 quai Ernest-Ansermet, CH-1211 Gen\`eve 4, Switzerland\label{aff81}
\and
Department of Physics, P.O. Box 64, University of Helsinki, 00014 Helsinki, Finland\label{aff82}
\and
Helsinki Institute of Physics, Gustaf H{\"a}llstr{\"o}min katu 2, University of Helsinki, 00014 Helsinki, Finland\label{aff83}
\and
Centre de Calcul de l'IN2P3/CNRS, 21 avenue Pierre de Coubertin 69627 Villeurbanne Cedex, France\label{aff84}
\and
Laboratoire d'etude de l'Univers et des phenomenes eXtremes, Observatoire de Paris, Universit\'e PSL, Sorbonne Universit\'e, CNRS, 92190 Meudon, France\label{aff85}
\and
SKAO, Jodrell Bank, Lower Withington, Macclesfield SK11 9FT, UK\label{aff86}
\and
University of Applied Sciences and Arts of Northwestern Switzerland, School of Computer Science, 5210 Windisch, Switzerland\label{aff87}
\and
Dipartimento di Fisica e Astronomia "Augusto Righi" - Alma Mater Studiorum Universit\`a di Bologna, via Piero Gobetti 93/2, 40129 Bologna, Italy\label{aff88}
\and
Infrared Processing and Analysis Center, California Institute of Technology, Pasadena, CA 91125, USA\label{aff89}
\and
Niels Bohr Institute, University of Copenhagen, Jagtvej 128, 2200 Copenhagen, Denmark\label{aff90}
\and
Universit\'e Paris Cit\'e, CNRS, Astroparticule et Cosmologie, 75013 Paris, France\label{aff91}
\and
CNRS-UCB International Research Laboratory, Centre Pierre Bin\'etruy, IRL2007, CPB-IN2P3, Berkeley, USA\label{aff92}
\and
University of Applied Sciences and Arts of Northwestern Switzerland, School of Engineering, 5210 Windisch, Switzerland\label{aff93}
\and
Institute of Physics, Laboratory of Astrophysics, Ecole Polytechnique F\'ed\'erale de Lausanne (EPFL), Observatoire de Sauverny, 1290 Versoix, Switzerland\label{aff94}
\and
Dipartimento di Fisica - Sezione di Astronomia, Universit\`a di Trieste, Via Tiepolo 11, 34131 Trieste, Italy\label{aff95}
\and
Telespazio UK S.L. for European Space Agency (ESA), Camino bajo del Castillo, s/n, Urbanizacion Villafranca del Castillo, Villanueva de la Ca\~nada, 28692 Madrid, Spain\label{aff96}
\and
Institut de F\'{i}sica d'Altes Energies (IFAE), The Barcelona Institute of Science and Technology, Campus UAB, 08193 Bellaterra (Barcelona), Spain\label{aff97}
\and
European Space Agency/ESTEC, Keplerlaan 1, 2201 AZ Noordwijk, The Netherlands\label{aff98}
\and
Waterloo Centre for Astrophysics, University of Waterloo, Waterloo, Ontario N2L 3G1, Canada\label{aff99}
\and
Department of Physics and Astronomy, University of Waterloo, Waterloo, Ontario N2L 3G1, Canada\label{aff100}
\and
Perimeter Institute for Theoretical Physics, Waterloo, Ontario N2L 2Y5, Canada\label{aff101}
\and
Space Science Data Center, Italian Space Agency, via del Politecnico snc, 00133 Roma, Italy\label{aff102}
\and
Institute of Space Science, Str. Atomistilor, nr. 409 M\u{a}gurele, Ilfov, 077125, Romania\label{aff103}
\and
Consejo Superior de Investigaciones Cientificas, Calle Serrano 117, 28006 Madrid, Spain\label{aff104}
\and
Universidad de La Laguna, Dpto. Astrof\'\i sica, E-38206 La Laguna, Tenerife, Spain\label{aff105}
\and
Institute for Particle Physics and Astrophysics, Dept. of Physics, ETH Zurich, Wolfgang-Pauli-Strasse 27, 8093 Zurich, Switzerland\label{aff106}
\and
Dipartimento di Fisica e Astronomia "G. Galilei", Universit\`a di Padova, Via Marzolo 8, 35131 Padova, Italy\label{aff107}
\and
INFN-Padova, Via Marzolo 8, 35131 Padova, Italy\label{aff108}
\and
Instituto de F\'isica Te\'orica UAM-CSIC, Campus de Cantoblanco, 28049 Madrid, Spain\label{aff109}
\and
Institut de Recherche en Astrophysique et Plan\'etologie (IRAP), Universit\'e de Toulouse, CNRS, UPS, CNES, 14 Av. Edouard Belin, 31400 Toulouse, France\label{aff110}
\and
Universit\'e St Joseph; Faculty of Sciences, Beirut, Lebanon\label{aff111}
\and
Departamento de F\'isica, FCFM, Universidad de Chile, Blanco Encalada 2008, Santiago, Chile\label{aff112}
\and
Universit\"at Innsbruck, Institut f\"ur Astro- und Teilchenphysik, Technikerstr. 25/8, 6020 Innsbruck, Austria\label{aff113}
\and
Department of Physics and Helsinki Institute of Physics, Gustaf H\"allstr\"omin katu 2, University of Helsinki, 00014 Helsinki, Finland\label{aff114}
\and
Instituto de Astrof\'isica e Ci\^encias do Espa\c{c}o, Faculdade de Ci\^encias, Universidade de Lisboa, Tapada da Ajuda, 1349-018 Lisboa, Portugal\label{aff115}
\and
Cosmic Dawn Center (DAWN)\label{aff116}
\and
Universidad Polit\'ecnica de Cartagena, Departamento de Electr\'onica y Tecnolog\'ia de Computadoras,  Plaza del Hospital 1, 30202 Cartagena, Spain\label{aff117}
\and
European University of Technology EUt+, European Union\label{aff118}
\and
Centre for Information Technology, University of Groningen, P.O. Box 11044, 9700 CA Groningen, The Netherlands\label{aff119}
\and
Dipartimento di Fisica e Scienze della Terra, Universit\`a degli Studi di Ferrara, Via Giuseppe Saragat 1, 44122 Ferrara, Italy\label{aff120}
\and
Istituto Nazionale di Fisica Nucleare, Sezione di Ferrara, Via Giuseppe Saragat 1, 44122 Ferrara, Italy\label{aff121}
\and
Universit\'e C\^{o}te d'Azur, Observatoire de la C\^{o}te d'Azur, CNRS, Laboratoire Lagrange, Bd de l'Observatoire, CS 34229, 06304 Nice cedex 4, France\label{aff122}
\and
ICSC - Centro Nazionale di Ricerca in High Performance Computing, Big Data e Quantum Computing, Via Magnanelli 2, Bologna, Italy\label{aff123}
\and
INAF, Istituto di Radioastronomia, Via Piero Gobetti 101, 40129 Bologna, Italy\label{aff124}
\and
Astronomical Observatory of the Autonomous Region of the Aosta Valley (OAVdA), Loc. Lignan 39, I-11020, Nus (Aosta Valley), Italy\label{aff125}
\and
Department of Mathematical Physics, Institute of Physics, University of S\~ao Paulo, R. do Mat\~ao 1371, 05508-090, S\~ao Paulo, SP, Brazil\label{aff126}
\and
CEA Saclay, DFR/IRFU, Service d'Astrophysique, Bat. 709, 91191 Gif-sur-Yvette, France\label{aff127}
\and
Code 665, NASA/GSFC, 8800 Greenbelt Road, Greenbelt, MD 20771, USA\label{aff128}
\and
Department of Astronomy, University of Maryland, College Park, MD 20742, USA\label{aff129}
\and
Oskar Klein Centre for Cosmoparticle Physics, Department of Physics, Stockholm University, Stockholm, SE-106 91, Sweden\label{aff130}
\and
Astrophysics Group, Blackett Laboratory, Imperial College London, London SW7 2AZ, UK\label{aff131}
\and
Univ. Grenoble Alpes, CNRS, Grenoble INP, LPSC-IN2P3, 53, Avenue des Martyrs, 38000 Grenoble, France\label{aff132}
\and
Dipartimento di Fisica, Sapienza Universit\`a di Roma, Piazzale Aldo Moro 2, 00185 Roma, Italy\label{aff133}
\and
Institut d'Astrophysique de Paris, 98bis Boulevard Arago, 75014, Paris, France\label{aff134}
\and
ICL, Junia, Universit\'e Catholique de Lille, LITL, 59000 Lille, France\label{aff135}
\and
Center for Astrophysics and Cosmology, University of Nova Gorica, Nova Gorica, Slovenia\label{aff136}
\and
MIT Kavli Institute for Astrophysics and Space Research, Massachusetts Institute of Technology, Cambridge, MA 02139, USA\label{aff137}
\and
CERCA/ISO, Department of Physics, Case Western Reserve University, 10900 Euclid Avenue, Cleveland, OH 44106, USA\label{aff138}
\and
Laboratoire Univers et Th\'eorie, Observatoire de Paris, Universit\'e PSL, Universit\'e Paris Cit\'e, CNRS, 92190 Meudon, France\label{aff139}
\and
Departamento de F{\'\i}sica Fundamental. Universidad de Salamanca. Plaza de la Merced s/n. 37008 Salamanca, Spain\label{aff140}
\and
Universiteit Gent, Department of Physics and Astronomy, Proeftuinstraat 86 N3, 9000 Ghent, Belgium
\label{aff141}
\and
Universit\'e de Strasbourg, CNRS, Observatoire astronomique de Strasbourg, UMR 7550, 67000 Strasbourg, France\label{aff142}
\and
Center for Data-Driven Discovery, Kavli IPMU (WPI), UTIAS, The University of Tokyo, Kashiwa, Chiba 277-8583, Japan\label{aff143}
\and
Scuola Superiore Meridionale, Via Mezzocannone 4, 80138, Napoli, Italy\label{aff144}
\and
California Institute of Technology, 1200 E California Blvd, Pasadena, CA 91125, USA\label{aff145}
\and
Departamento de F\'{i}sica Te\'{o}rica, At\'{o}mica y \'{O}ptica, Universidad de Valladolid, 47011 Valladolid, Spain\label{aff146}
\and
Laboratory for Disruptive Interdisciplinary Science (LaDIS), Universidad de Valladolid, 47011 Valladolid, Spain\label{aff147}
\and
Department of Physics, Astronomy and Mathematics, University of Hertfordshire, College Lane, Hatfield AL10 9AB, UK\label{aff148}
\and
Institute of Cosmology and Gravitation, University of Portsmouth, Portsmouth PO1 3FX, UK\label{aff149}
\and
Centre de Recherche Astrophysique de Lyon, UMR5574, CNRS, Universit\'e Claude Bernard Lyon 1, ENS de Lyon, 69230, Saint-Genis-Laval, France\label{aff150}
\and
Department of Physics \& Astronomy, University of California Irvine, Irvine CA 92697, USA\label{aff151}
\and
Departamento F\'isica Aplicada, Universidad Polit\'ecnica de Cartagena, Campus Muralla del Mar, 30202 Cartagena, Murcia, Spain\label{aff152}
\and
Instituto de F\'isica de Cantabria, Edificio Juan Jord\'a, Avenida de los Castros, 39005 Santander, Spain\label{aff153}
\and
Observatorio Nacional, Rua General Jose Cristino, 77-Bairro Imperial de Sao Cristovao, Rio de Janeiro, 20921-400, Brazil\label{aff154}
\and
Department of Mathematics and Physics E. De Giorgi, University of Salento, Via per Arnesano, CP-I93, 73100, Lecce, Italy\label{aff155}
\and
INFN, Sezione di Lecce, Via per Arnesano, CP-193, 73100, Lecce, Italy\label{aff156}
\and
INAF-Sezione di Lecce, c/o Dipartimento Matematica e Fisica, Via per Arnesano, 73100, Lecce, Italy\label{aff157}
\and
Department of Computer Science, Aalto University, PO Box 15400, Espoo, FI-00 076, Finland\label{aff158}
\and
School of Physics and Astronomy, University of Nottingham, University Park, Nottingham NG7 2RD, UK\label{aff159}
\and
Ruhr University Bochum, Faculty of Physics and Astronomy, Astronomical Institute (AIRUB), German Centre for Cosmological Lensing (GCCL), 44780 Bochum, Germany\label{aff160}
\and
DARK, Niels Bohr Institute, University of Copenhagen, Jagtvej 155, 2200 Copenhagen, Denmark\label{aff161}
\and
INAF-Osservatorio Astrofisico di Arcetri, Largo E. Fermi 5, 50125, Firenze, Italy\label{aff162}
\and
Astrophysics Research Centre, University of KwaZulu-Natal, Westville Campus, Durban 4041, South Africa\label{aff163}
\and
School of Mathematics, Statistics \& Computer Science, University of KwaZulu-Natal, Westville Campus, Durban 4041, South Africa\label{aff164}
\and
Department of Physics and Astronomy, Vesilinnantie 5, University of Turku, 20014 Turku, Finland\label{aff165}
\and
Finnish Centre for Astronomy with ESO (FINCA), Quantum, Vesilinnantie 5, University of Turku, 20014 Turku, Finland\label{aff166}
\and
Serco for European Space Agency (ESA), Camino bajo del Castillo, s/n, Urbanizacion Villafranca del Castillo, Villanueva de la Ca\~nada, 28692 Madrid, Spain\label{aff167}
\and
ARC Centre of Excellence for Dark Matter Particle Physics, Melbourne, Australia\label{aff168}
\and
Centre for Astrophysics \& Supercomputing, Swinburne University of Technology,  Hawthorn, Victoria 3122, Australia\label{aff169}
\and
Department of Physics and Astronomy, University of the Western Cape, Bellville, Cape Town, 7535, South Africa\label{aff170}
\and
MTA-CSFK Lend\"ulet Large-Scale Structure Research Group, Konkoly-Thege Mikl\'os \'ut 15-17, H-1121 Budapest, Hungary\label{aff171}
\and
Konkoly Observatory, HUN-REN CSFK, MTA Centre of Excellence, Budapest, Konkoly Thege Mikl\'os {\'u}t 15-17. H-1121, Hungary\label{aff172}
\and
Ohio University, Physics \& Astronomy Department,1 Ohio University, Athens, OH 45701, USA\label{aff173}
\and
Brazilian Center for Research in Physics (CBPF), Dr. Xavier Sigaud st. 150, zip 22290-180, Rio de Janeiro, RJ, Brazil\label{aff174}
\and
Institute for Theoretical Particle Physics and Cosmology (TTK), RWTH Aachen University, 52056 Aachen, Germany\label{aff175}
\and
IRFU, CEA, Universit\'e Paris-Saclay 91191 Gif-sur-Yvette Cedex, France\label{aff176}
\and
Centro de Astrof\'{\i}sica da Universidade do Porto, Rua das Estrelas, 4150-762 Porto, Portugal\label{aff177}
\and
Dipartimento di Fisica, Universit\`a di Roma Tor Vergata, Via della Ricerca Scientifica 1, Roma, Italy\label{aff178}
\and
INFN, Sezione di Roma 2, Via della Ricerca Scientifica 1, Roma, Italy\label{aff179}
\and
HE Space for European Space Agency (ESA), Camino bajo del Castillo, s/n, Urbanizacion Villafranca del Castillo, Villanueva de la Ca\~nada, 28692 Madrid, Spain\label{aff180}
\and
STAR Institute, University of Li{\`e}ge, Quartier Agora, All\'ee du six Ao\^ut 19c, 4000 Li\`ege, Belgium\label{aff181}
\and
Dipartimento di Fisica ``G. Occhialini", Universit\`a degli Studi di Milano Bicocca, Piazza della Scienza 3, 20126 Milano, Italy\label{aff182}
\and
School of Physical Sciences, The Open University, Milton Keynes, MK7 6AA, UK\label{aff183}
\and
Department of Astrophysics, University of Zurich, Winterthurerstrasse 190, 8057 Zurich, Switzerland\label{aff184}
\and
Space physics and astronomy research unit, University of Oulu, Pentti Kaiteran katu 1, FI-90014 Oulu, Finland\label{aff185}
\and
Department of Mathematics and Physics, Roma Tre University, Via della Vasca Navale 84, 00146 Rome, Italy\label{aff186}
\and
Theoretical astrophysics, Department of Physics and Astronomy, Uppsala University, Box 524, 751 20 Uppsala, Sweden\label{aff187}
\and
Mathematical Institute, University of Leiden, Einsteinweg 55, 2333 CA Leiden, The Netherlands\label{aff188}
\and
School of Physics \& Astronomy, University of Southampton, Highfield Campus, Southampton SO17 1BJ, UK\label{aff189}
\and
Kobayashi-Maskawa Institute for the Origin of Particles and the Universe, Nagoya University, Chikusa-ku, Nagoya, 464-8602, Japan\label{aff190}
\and
Department of Astrophysical Sciences, Peyton Hall, Princeton University, Princeton, NJ 08544, USA\label{aff191}
\and
Fakult\"at f\"ur Physik, Universit\"at Bielefeld, Postfach 100131, 33501 Bielefeld, Germany\label{aff192}
\and
School of Mathematics, Statistics and Physics, Newcastle University, Herschel Building, Newcastle-upon-Tyne, NE1 7RU, UK\label{aff193}
\and
SRON Netherlands Institute for Space Research, Landleven 12, 9747 AD, Groningen, The Netherlands\label{aff194}
\and
Department of Physics and Astronomy, University of British Columbia, Vancouver, BC V6T 1Z1, Canada\label{aff195}}    

\date\today

%
%

\abstract{This paper describes the characteristics of the \Euclid visible imager (VIS) data delivered as part of the Euclid \ac{DR1}. This data set contains 42\,092 calibrated frames. Validated nominal exposures cover 2092~$\mathrm{deg}^2$ of sky. We outline the minor changes made to the VIS \ac{PF} since previous releases. For the wide survey, the photometric and astrometric accuracy of the reference stellar sample ($18.5<\IE<19.8$) is excellent. The internal and external astrometric dispersions are $0.08$ and $0.18\,\mathrm{mas}$, respectively, at the quadrant level, and $3.4$ and $5.6\,\mathrm{mas}$ at the source level. The corresponding internal and external photometric dispersions are $0.76$ and $1.4\,\mathrm{mmag}$ at the quadrant level, and $11$ and $25\,\mathrm{mmag}$ at the source level. We characterise remaining astrometric and photometric systematic effects: focal-plane-dependent variations; time-dependent photometric offsets; and colour-dependent centroid shifts. The noise floor in the deepest VIS images appears to be set by the extended diffuse light of faint galaxies. Globally, the \ac{DR1} VIS data meet pre-launch mission requirements and the 42\,092 images it contains provide our first substantial look at the sky as seen by VIS.   
}


   \titlerunning{DR1: VIS processing and data products}
   \authorrunning{Euclid Collaboration: H.~J.~McCracken et al.}
   
   \maketitle  \acresetall 
%
%
%
%
   
\section{\label{sc:Intro}Introduction}

The ESA \Euclid mission aims to improve our understanding of dark energy by precisely measuring the shapes and positions of an unprecedentedly large sample of galaxies \citep{EuclidSkyOverview}. A key element of \Euclid is the \ac{VIS} camera \citep{EuclidSkyVIS}. This paper describes the processing and characteristics of the VIS images delivered as part of the Euclid \acl{DR1}~\citep[\acs{DR1};][]{DR1cite,DR1-TP001}\acused{DR1}, which were processed by the \Euclid \acl{SGS}~\citep[\acs{SGS};][]{DR1-TP033}\acused{SGS}. Previously, \acl{Q1}~\citep[\acs{Q1};][hereafter \citetalias{Q1-TP002}]{Q1-TP002}\acused{Q1} and \acl{Q2}~\citep[\acs{Q2};][]{Q2-TP002}\acused{Q2} provided $61.3$ and $4.8\,\mathrm{deg}^2$ of VIS data, respectively. DR1 data contain nominal VIS exposures over 2092.2~$\mathrm{deg}^2$ spanning the \ac{EWS} and \ac{EDS} areas, and also include the data in the COSMOS area and other calibration fields. \ac{DR1} comprises principally survey data taken from 14 February 2024 to 21 May 2025. It also includes calibration fields and observations taken before the survey started during the \ac{PV} phase. This paper presents a description of the characteristics of the VIS DR1 data set, an assessment of the performance of the \ac{VIS} \ac{PF} pipeline, and a detailed discussion of the global properties of the processed VIS \ac{DR1} data.  

The plan of this paper is as follows: \cref{sc:PipeOver} summarises the \ac{VIS} \ac{PF} and outlines the differences compared to \ac{Q1} processing; \cref{sc:DR1Perf} explains how \ac{DR1} input data were validated; \cref{sc:DR1GlobalPerf} presents an investigation of the global performance of all the VIS images in \ac{DR1} and  \cref{sc:caveats} outlines some remaining problems. Conclusions are presented in \cref{sc:Conc}.

\section{\label{sc:PipeOver} Processing VIS DR1 data}

\subsection{\label{sc:brdesc} A brief summary of the DR1 VIS processing function}

VIS is an array of 36 back-illuminated $40\,\micron$-thick \acp{CCD}, each $4132\times4096$ pixels in size. Each square pixel is $12\,\micron$ on a side. At the centre of the \ac{FPA} the pixel scale is \ang{;;0.100} and it varies by 1.5\% across it, well within the required \ang{;;0.10}\,$\pm$\,\ang{;;0.01} (MRD-WL-003 in \citealt{DR1-TP026}). Each CCD has four quadrants, each of which has an independent set of readout electronics. The basic unit of \Euclid observations is the \acl{ROS}~\citep[\acs{ROS};][]{DR1-TP027}\acused{ROS}. Each \ac{ROS} consists of a nominal 566\,s exposure at each of four dithers in an `S'-pattern, with an additional two short science 95\,s exposures at the first two positions of the sequence. Observations are grouped together in `patches' or `skypatches' defined in \cite{Scaramella-EP1} as `a compact set of fields with a valid observation sequence'. 

After transmission from the satellite, `Level 1' software running at the \ac{SOC} assembles these compressed raw data into 144-extension \ac{FITS} files with a basic set of header keywords. In the next step, data are processed by the \ac{VIS} \ac{PF}, grouped by `observation id' or \texttt{obs\_id}. An \texttt{obs\_id} may contain more than one \ac{ROS}. 

A complete description of each \ac{VIS} \ac{PF} processing step is provided in \citetalias{Q1-TP002}. In DR1, there are no major algorithmic changes compared to \ac{Q1} or \ac{Q2} processing. A brief description of the small changes can be found in \cref{sc:DR1changes}. Version 18.0.4 of \ac{VIS} \ac{PF} was used to process DR1, and its source code is made available as part of this release. 

The two main functional units of the \ac{VIS} \ac{PF} are a set of 16 `calibration pipelines' that compute master calibration products for each instrumental effect, and a `science pipeline' that uses these calibration products to correct and process the raw data. Many of the calibrations described below are derived from monthly visits of the \Euclid self-calibration field, described in section 4.2.3 of \citet{EuclidSkyOverview} and section 5 in \cite{DR1-TP026}. Each self-calibration visit is made at a different instrument position angle, which is useful for determining whether residual systematic effects are present at the sky-frame level or instrument-frame level (\cref{sc:selfcal}). 

The science pipeline also performs quality assessment, which enables the grading of the output products, an essential step to ensure the scientific quality of the final products. In \citetalias{Q1-TP002} we divided the science pipeline into three main blocks, which are summarised below. Each block comprises several \acp{PE}. The functionality of each PE is limited to one or two functions. 

The first set of \acp{PE} reads raw VIS exposures, flags saturated pixels, applies a crosstalk correction, bias correction, and non-linearity correction. This is followed by cosmic-ray flagging using \texttt{LACosmic} \citep{2001PASP..113.1420V}, \ac{BFE} correction, \ac{PRNU} correction, and gain equalisation. The input calibration products in this step are the cosmetic maps, crosstalk model, master bias, non-linearity model, \ac{BFE} model, master \ac{PRNU}, and the gain model. This block outputs flag maps indicating the location of bad pixels together with corrected images. \Ac{CTI}, the smearing of charge during readout caused by defects in the detector silicon lattice (`traps'), is not corrected in DR1. However, we are monitoring the trap density in VIS and may activate this correction in future releases.  

The second set of \acp{PE} takes these corrected images, applies an illumination correction, estimates the background (which is supplied as a data product but not subtracted from the images), extracts an intermediate source catalogue using \texttt{SExtractor} \citep{1996A&AS..117..393B}, computes an astrometric solution, flags ghosts, computes a second intermediate photometric catalogue, and uses this photometric catalogue to perform a photometric calibration. The input products to this block comprise the illumination model, the \Gaia catalogue, an astrometric distortion model, the ghost model, the \ac{PSF} model computed using \texttt{PSFex} \citep{2011ASPC..442..435B}, and the zero-point reference photometry. The output products include a background map computed using \texttt{NoiseChisel} \citep{2015ApJS..220....1A}, the pointing reconstruction, an updated flag map, and  astrometrically and photometrically calibrated single-exposure images together with a calibrated source catalogue. Individual exposures are not background subtracted. 

In the last step, the processing block takes these calibrated images and, for each \ac{ROS}, stacks them together using \texttt{SWarp} \citep{2002ASPC..281..228B} after subtracting the supplied \texttt{NoiseChisel} background from each exposure. Catalogues are extracted from these stacked images using \texttt{SExtractor}, which are then used to compute a \ac{PSF} model and an associated catalogue. 

In the \fnurl{\Euclid Science Archive,}{https://eas.esac.esa.int/sas/} VIS products are delivered as \texttt{DpdVisCalibratedQuadFrame} (calibrated frame, background map, weight map and \ac{PSF} model),  \texttt{DpdVisCalibratedFrameCatalog} (single-exposure catalogues), and \texttt{DpdVisStackedFrame}, and \texttt{DpdVisStackedFrameCatalog} (the stacks and their associated catalogues). The unprocessed input to VIS \ac{PF}, \texttt{DpdVisRawFrame}, is also distributed. 

The format and content of these products are described in the \fnurl{\ac{DR1} \ac{DPDD}.}{https://euclid.esac.esa.int/dr/dr1/dpdd/} The \fnurl{\ac{DR1} Explanatory Supplement}{https://euclid.esac.esa.int/dr/dr1/expsup/} provides a complete list of known processing issues with \ac{VIS} \ac{DR1} data.  Because \texttt{DATE-OBS} does not mark the start of the integration, a recipe for computing the exact time of the shutter opening is given in Appendix~\ref{sc:shutter-time}.

\subsection{A brief description of DR1 VIS calibration }

\cite{DR1-TP026} provides a complete account of the \Euclid calibration strategy. Since \ac{DR1} contains data taken over 15.2 months, a careful accounting of the validity of each calibration product is required. In the \ac{EWS}, there are 50 skypatches (although only 49 are used for the astrometric measurements). 

We note that DR1 is a reprocessing of data already reduced by the \ac{SGS} in the \ac{OTF} campaigns. The \ac{OTF} pipeline is executed on all VIS science data a few days after acquisition using a set of standard calibration products. Analysing these data allows an assessment of the validity of the raw data and corresponding calibration products for the release reprocessing. \ac{OTF} reprocessing also enables the validation of \ac{VIS} \ac{PF} algorithms with a wide variety of observations taken under different observing conditions. 

For DR1, a single calibration product for the entire period was used for gain, \ac{PSF} model, readout noise, saturation, \ac{BFE}, ghost masking, crosstalk, and stellar mask corrections. For the flag maps, master biases, \ac{PRNU}, astrometric distortion, and zero-point model, a series of products were computed with different validity periods, as explained below. 

Analysing \ac{OTF} data processed with the same astrometric distortion model showed steadily increasing astrometric residuals compared to the \Gaia reference catalogues. Therefore, for \ac{DR1}, seven different distortion models were computed from seven separate self-calibration visits. 

\begin{figure}[htbp]
    \centering
    \includegraphics[width=\hsize]{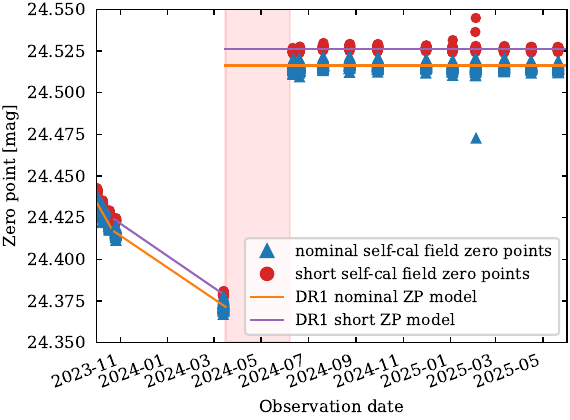}
    \hfill
    \caption{Evolution of the VIS DR1 photometric zero point. Symbols show per-visit \Gaia-derived zero points from the self-calibration field. The lines show the two-segment linear model used to compute a zero point for each DR1 exposure. The gradual decline is mainly driven by ice-induced throughput loss, with a step at a decontamination event. The difference between the nominal and short exposures is due to an error in aperture corrections, fixed in later versions of the pipeline. In the shaded region, the VIS instrument throughput was heavily degraded by ice contamination, and these data are not part of DR1.}
    \label{fig:zeropoint-evolution}
\end{figure}

Initially, the VIS photometric zero point was determined at each self-calibration visit from observations of \Gaia stars in the self-calibration field. However, until March 2024, these zero points lost their validity after approximately one week primarily because of ice contamination on a folding mirror and the resulting scatter loss \citep[sections 5 and 6 in][]{DR1-TP026b}. Therefore, using all the self-calibrations in DR1, we derived a zero point for each self-calibration observation and fitted all these observations with a two-segment linear model, with a constant after the decontamination event. This model was then used to compute an individual zero point for each image. \Cref{fig:zeropoint-evolution} shows the evolution of this zero point over the period covered by DR1 observations. The small zero-point differences between nominal and short exposures are a consequence of an error in how aperture corrections were computed for both types of exposures which has since been corrected. 

\subsection{\label{sc:DR1changes}Changes compared to Q1}

The main changes between the Q1 and DR1 versions of \ac{VIS} \ac{PF} are minor. They concern cosmic-ray rejection, photometric calibration, background estimation, stack production, and the content of the delivered data products. In DR1, cosmic-ray identification includes a second detection pass that uses the stacked image to identify cosmic rays. The photometric calibration has been updated through a revised filtering of \Gaia reference sources and by adopting the \Gaia-to-\IE\ transformation from the mission database (Appendix~\ref{sc:gaia-ie}).

Several changes were made in the image combination stage. The background is now subtracted from the individual exposures before combination. In addition, the stacked-frame data product has been simplified: the background FITS file is no longer included in \texttt{DpdVisStackedFrame}, and the \texttt{RMS} and \texttt{FLG} extensions have been removed from the stack FITS file. The VIS flag definition has also been updated by removing the flag name \texttt{GOOD}; the corresponding numerical flag values are unchanged.

The content of the calibrated frame products has been extended. New columns were added to the calibrated frame catalogues, and the \ac{VIS} \ac{PSF} model is now included in \texttt{DpdVisCalibratedQuadFrame}. Finally, FITS headers of the extensions now contain the keywords \texttt{MJD-OBS} and \texttt{TIMESYS=`UTC'}.

\section{\label{sc:DR1Perf} DR1 input data validation}

After the \ac{VIS} \ac{PF} is executed, reports with \ac {DQC} statistics are generated for each image in each skypatch as described in section 7 of \citetalias{Q1-TP002}. These parameters include the combined image `score' together with statistics indicating the number of cosmic-ray hits, the astrometric residuals, and the image pointing stability measured from the \texttt{spread\_model} parameter. These image quality statistics, derived from the \ac{OTF} processing, are used to identify which images should be rejected from the \ac{DR1} reprocessing. The most common reason for image rejection is excessive cosmic-ray hits due to solar activity. Furthermore, the VIS data products in skypatches 39 and 40 should be used with caution because their astrometric solutions were processed with an inappropriate distortion model (see \cref{sc:astrpatch3940} for more details). Quality control statistics are supplied as \texttt{json} and \texttt{xml} files with the release. They are plotted for every exposure in the Explanatory Supplement, which also provides a complete list of rejected frames.   

\begin{figure}[htbp]
\centering
\includegraphics[width=\hsize]{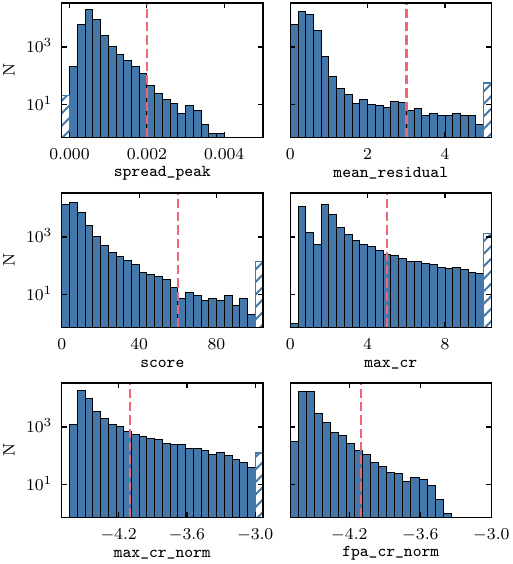}
\caption{Distributions of the six image-level \ac{DQC} statistics for the \num{42092} VIS images in DR1. \emph{Top row}: the \texttt{spread\_peak}
image-quality parameter and the mean astrometric residual \texttt{mean\_residual}. \emph{Middle row}: the image \texttt{score} and the peak cosmic-ray statistic \texttt{max\_cr}. \emph{Bottom row}: the normalised cosmic-ray statistics \texttt{max\_cr\_norm} and \texttt{fpa\_cr\_norm}. Red dashed lines indicate thresholds. A warning is raised if any statistic exceeds its threshold; after applying these quality-control thresholds \num{37875} images remain and \num{4217} were flagged. Nevertheless, all \num{42092} images were delivered as part of DR1. The hatched overflow bars indicate the total number of observations beyond the axis range.} \label{fig:dr1-dqc}
\end{figure}

\Cref{fig:dr1-dqc} shows the distribution of image quality control statistics for all VIS images in DR1. If any of these quality control thresholds is exceeded, a warning is raised and the image should be used with caution (for example, one quadrant has a higher-than-normal amount of cosmic rays).  

\section{\label{sc:DR1GlobalPerf} Global performance analysis of VIS DR1 data}

In \Euclid, for the cosmological tracer population, tiny systematic errors are the dominant source of uncertainty. Given that \ac{VIS} \ac{PF} is one of the first \acp{PF} in the processing chain, it is important to assess the performance of the entire VIS DR1 data set to ensure that any systematic effects are properly quantified. 

\subsection{\label{sc:CatCreation} Creation of the DR1 merged catalogues and reference sample}

We create matched catalogues for all observations in DR1 from the single-exposure (and stacked) catalogues produced by \ac{VIS} \ac{PF}. Because the DR1 data volume is large, creation of a single monolithic catalogue is impractical. Therefore, matched catalogues are created for each skypatch following a similar procedure to the one outlined in \citetalias{Q1-TP002}. From the individual VIS calibrated frame catalogues, two catalogues are produced: one matched with \Gaia DR3 \citep{2023A&A...674A...1G} and containing only point-like  sources, and one containing both point-like and extended sources. 

For the point-source catalogue, unresolved sources are selected using two
criteria. Firstly, a cut on the radius enclosing $50\%$ of the flux, $1.0 <
\texttt{FLUX\_RADIUS} < 1.3$, is designed to select star-like sources. Secondly, we
apply an exposure-time-dependent criterion based on the 50-pixel diameter
aperture magnitude \texttt{MAG\_APER}: for nominal exposures ($t_{\mathrm{exp}} > 500$\,s), sources are selected with $18.5 < \texttt{MAG\_APER} < 22.0$, while for short exposures ($t_{\mathrm{exp}} < 100$\,s), the range is $16.5 < \texttt{MAG\_APER} < 21.0$. The bright limits are chosen to minimise the number of saturated pixels in each source.  

These VIS detections are cross-matched with \Gaia DR3 sources within $\ang{;;0.5}$. The positions of the matched \Gaia sources are then propagated to the VIS observation epoch using proper motions, corrected for parallax as described below, and astrometric residuals are computed for each source.

The \ac{VIS} \ac{PF} computes the \ac{WCS} of each exposure using \Gaia DR3 positions transformed to the apparent directions seen from \Euclid at the time of observation, including parallax. The fitted \ac{WCS} is then converted back to the \ac{ICRS}. \ac{WCS} transformations map pixel coordinates to sky directions without requiring source distances, so this conversion removes the distance-independent effects of aberration and light deflection but retains parallax, which depends on source distance. VIS catalogue positions therefore still contain the parallactic displacement of each source as seen from \Euclid. To compare with \Gaia, we add to each \Gaia position the parallactic displacement for an observer at \Euclid,
\begin{equation}
\begin{aligned}
\Delta\alpha_\varpi\cos\delta &= \varpi\,(b_x\sin\alpha - b_y\cos\alpha)\,,\\
\Delta\delta_\varpi &= \varpi\,(b_x\cos\alpha\sin\delta + b_y\sin\alpha\sin\delta - b_z\cos\delta)\,,
\end{aligned}
\label{eq:plx}
\end{equation}
where $(\Delta\alpha_\varpi\cos\delta, \Delta\delta_\varpi)$ is the displacement of a source at right ascension $\alpha$ and declination $\delta$ from its barycentric to its apparent position, $\varpi$ is its measured \Gaia parallax (negative values included), and $(b_x, b_y, b_z)$ is the barycentric position of \Euclid\ in au, in equatorial coordinates with $x$ towards $(\alpha, \delta) = (0, 0)$ and $z$ towards the north celestial pole; the displacement then has the units of $\varpi$. We approximate the position of \Euclid\ as that of the Earth displaced 0.01\,au further away from the Sun. Because \Euclid\ observes at a solar aspect angle close to $90\degr$ (median $91\degr$), with its roll angle fixed relative to the direction of the Sun, this displacement lies almost along $+X_{\rm FPA}$ in every exposure, with an amplitude close to $\varpi$. For our reference sample the mean displacement is $0.73\,\mathrm{mas}$; without this correction, it would appear as a coherent offset in $X_{\rm FPA}$ across the focal plane.

As a validation of our cross-match method, we considered a representative subset of 98 nominal exposures (two per skypatch, over 49 skypatches). Here,  $84.0\%$ of selected point sources with $18.5 < \IE < 19.8$ are matched with \Gaia; the unmatched sources are mostly cosmic-ray hits, flagged in the VIS flag map, that pass our simple selection criteria. To test the proper-motion propagation, we also matched \num{65848} isolated \Gaia stars with $18.5 < G < 20$ lying at least 20 pixels inside a quadrant: $1.5\%$ have no VIS counterpart. Half of these ($0.75\%$ of the stars, median proper motion $72\,\mathrm{mas\,yr^{-1}}$) are missed because the match is made at the \Gaia epoch 2016.0, before the proper motions are applied, and they therefore lie outside the \ang{;;0.5} match radius; the remaining $0.77\%$ are preferentially found near bright stars. Including these sources changes the mean residual by $0.018$\,mas and any quadrant mean by at most $0.12$\,mas. For DR2, residuals will be computed in the \ac{VIS} \ac{PF} pipeline by propagating \Gaia sources to the observation epoch before matching.

Source positions in the physical ($X_{\rm FPA}$, $Y_{\rm FPA}$) focal-plane coordinate system are given in millimetres, together with the quadrant ID for each source, and observation metadata such as position angle, skypatch identification, and exposure time. Physical coordinates are derived from the windowed pixel centroids of each source using a per-quadrant linear transformation derived from pre-launch metrology. 

Focal-plane-level statistics are also computed. For each exposure, a single catalogue is produced, combining the 144 quadrant extensions. These exposure-level catalogues are then grouped by skypatch and all repeated observations of the same source are given a unique identifier, \texttt{group\_id}. 

Predicted $\IE$ magnitudes are computed for \Gaia sources using the transformation described in Appendix~\ref{sc:gaia-ie}. In the last step, a reference sample for the nominal exposures is created by applying a further $18.5<\IE<19.8$ cut on the measured aperture-corrected VIS magnitudes and a \Gaia astrometric quality cut selecting only sources with $\mathrm{RUWE}<1.2$ (RUWE, the renormalised unit weight error, is an astrometric goodness-of-fit statistic) and requiring the proper-motion errors $\sigma_{\mu_{\alpha^*}}$ and $\sigma_{\mu_\delta}$ to both be less than $0.5\,\mathrm{mas\,yr^{-1}}$.

\begin{table*}[htbp]
\caption{%
  Internal and external astrometric and photometric accuracy of the
  VIS Euclid DR1 \ac{EWS} calibrated-frame catalogues in the $18.5<\IE<19.8$ reference sample
  after \Gaia\ astrometric quality cuts ($\mathrm{RUWE}<1.2$,
  $\sigma_{\mu_{\alpha^*}},\sigma_{\mu_\delta}<0.5\,\mathrm{mas\,yr^{-1}}$);
  $6\,845\,677$ astrometric measurements over $49$ skypatches and
  $6\,477\,228$ photometric measurements over $50$ skypatches (skypatch~40 is excluded from the astrometric analysis). The photometric sample excludes clipped measurements as described in \cref{sc:PhotAccInfo}. For the internal columns, the source-level $\sigma_{\rm s}$ is the single-measurement repeatability, corrected for the contribution of each measurement to the mean of its source. The source-level rows combine all measurements via the law of total
  variance, so $\sigma_{\rm s}$ captures the full source-level
  dispersion (shot noise, \Gaia\ proper-motion residuals, \ac{CCD}-scale
  scatter). The quadrant-level rows give the mean and dispersion
  of the $144$ per-quadrant cell values (\cref{eq:mean} for astrometry) shown in
  \cref{fig:int-external-astrom,fig:phot-internal-external-residuals},
  so $\sigma_{\rm q}$ reflects mainly \ac{FPA}-level systematics. The   patch-level rows give the mean and dispersion of the $49$   (astrometry) or $50$ (photometry) per-skypatch values, so   $\sigma_{\rm p}$ captures patch-to-patch systematics (for example, zero-point  drift between observation epochs or stellar-population differences).%
}
\label{tab:dr1-internal-external}
\centering
\begin{tabular}{l l S[table-format=-1.4] S[table-format=2.3] S[table-format=1.2] S[table-format=2.2]}
\hline\hline
\noalign{\smallskip}
Quantity & Aggregation & \multicolumn{2}{c}{Internal} & \multicolumn{2}{c}{External} \\
         &             & {$\mu$} & {$\sigma$} & {$\mu$} & {$\sigma$} \\
\hline
\noalign{\smallskip}
\multirow{3}{*}{Astrometry [mas]}
    & source-level  ($\sigma_{\rm s}$)   &  0.0007 & 3.4   & 0.31 & 5.6  \\
    & quadrant-level ($\sigma_{\rm q}$)&  0.14   & 0.08  & 0.37 & 0.18 \\
    & patch-level    ($\sigma_{\rm p}$)&  0.005  & 0.004 & 0.31 & 0.06 \\
\noalign{\smallskip}
\multirow{3}{*}{Photometry [mmag]}
    & source-level   ($\sigma_{\rm s}$)  & -0.5 & 11   & 3.8 & 25   \\
    & quadrant-level ($\sigma_{\rm q}$) & -0.5 & 0.76 & 3.8 & 1.4  \\
    & patch-level    ($\sigma_{\rm p}$)& -0.5 & 0.07 & 3.8 & 2.3  \\
\noalign{\smallskip}
\hline
\end{tabular}
\tablefoot{%
  External photometric residuals are $\IE-G'$  (\cref{sc:PhotAccInfo})  magnitudes, and internal ones are differences from the mean VIS magnitude of each source in mmag. Source-level dispersion combines the within-quadrant scatter with the  variation of the quadrant means (from the law of total variance).   Quadrant-level aggregation is the weighted standard deviation of the   $144$ per-quadrant cell means displayed in   \cref{fig:int-external-astrom,fig:phot-internal-external-residuals}.  Patch-level aggregation is the weighted standard deviation of the per-skypatch means in the reference bin.
}
\end{table*}

\subsection{\label{sc:AstAcc} Astrometric accuracy of the merged DR1 catalogues}

\subsubsection{\label{sc:AstAccGeo} Average focal-plane residual map}

\begin{figure*}[htbp]
    \centering
    \includegraphics[width=1\hsize]{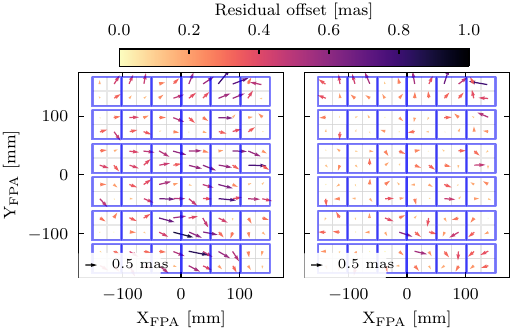}
    \hfill
    \caption{\emph{Left panel}: Average residuals per quadrant over 49 VIS DR1 \ac{EWS} skypatches for sources with  $18.5<\IE<19.8$, computed as VIS $-$ \Gaia  for each pair rotated into the VIS focal plane frame using the position angle of each observation and averaged over all observations. \emph{Right panel}: residuals after subtracting the weighted mean offset $\Delta X_{\rm FPA} = 0.30$, $\Delta Y_{\rm FPA} = -0.04$\,mas from each quadrant.}
    \label{fig:wide_astro_field_comparison}
\end{figure*}

For weak lensing, the geometric distortion on arcminute scales must be accurately known. The \Euclid requirements (MRD-WL-009 and R-CAL-A-VS-3000, Tables A.1 and A.2 in \citealt{DR1-TP026}) state that the \ac{RMS} post-calibration geometric distortion averaged within a circular aperture of radius \ang{;4;} should be smaller than $0.003\%$, which corresponds to $3\times10^{-5}$. Flowed down to the \ac{VIS} \ac{PF}, this requirement implies that the mean \ac{RMS} astrometric residual over all \Gaia reference stars in each exposure should be less than 3\,mas. 

We also test the parent requirement directly. For each of the 16\,702 nominal exposures in 49 skypatches (skypatch 40 is excluded; \cref{sc:astrpatch3940}), we average the signed VIS$-$\Gaia\ residuals within \ang{;4;}-radius apertures and difference these averages over \ang{;4;} to obtain the displacement gradient. After subtracting the measurement noise, estimated by splitting the stars in each exposure into two independent halves, the \ac{RMS} gradient over all exposures is $(7.0\pm0.3)\times10^{-7}$, and the largest value in any exposure is $(2.04\pm0.08)\times10^{-5}$, below the $3\times10^{-5}$ requirement.

For point-like sources with $18.5<\IE<19.8$ (chosen to match the object selection for photometric calibration) in \ac{EWS} nominal exposures, we computed the difference between  VIS source positions and the corresponding \Gaia DR3 positions (propagated to the VIS observation epoch, as above),  
\begin{equation}
      \Delta\alpha = (\alpha_{\rm VIS} -  \alpha_{\rm \textit{Gaia}})\,\cos\delta , \quad
      \Delta\delta = \delta_{\rm VIS} -  \delta_{\rm \textit{Gaia}} ,
\end{equation}
where $\cos\delta$ converts the difference in right ascension into an angle on the sky.
We then rotate these  sky-frame residuals into detector coordinates using the \ac{PA} of each observation,
  \begin{equation}
      \begin{pmatrix}
          \Delta X \\ \Delta Y
      \end{pmatrix}
      =
      \begin{pmatrix}
          \cos\theta & \sin\theta \\
          -\sin\theta & \cos\theta
      \end{pmatrix}
      \begin{pmatrix}
          -\Delta\alpha \\ \Delta\delta
      \end{pmatrix}\,,
      \label{eq:xy}
  \end{equation}
where $\theta = {\rm PA} + 90\degr$ accounts for the VIS coordinate system convention, and the minus sign accounts for the opposite handedness of the focal-plane frame with respect to the (east, north) sky frame. The final $(\Delta X, \Delta Y)$ represent residuals in the focal plane frame, measured in mas. The source positions are mapped to millimetres using the per-quadrant pixel-to-millimetre \ac{WCS} defined at the detector merging step.

To compute the inter-field averaged focal plane residual map, we first binned residuals by quadrant. Within each skypatch $p$ and quadrant $q$, we computed the mean residuals $(\mu_{x,pq}, \mu_{y,pq})$ and their standard errors. To combine measurements across the 49 DR1 skypatches, we applied inverse-variance weighting for each quadrant.  

These results are summarised in \cref{fig:wide_astro_field_comparison}. The arrows represent these inverse-variance weighted means $(\langle \mu_x \rangle_q,
\langle \mu_y \rangle_q)$ at each quadrant, with arrow colours indicating the residual magnitude $\sqrt{\langle \mu_x \rangle_q^2 + \langle \mu_y \rangle_q^2}$. The left panel shows the average residual per quadrant computed on the DR1 \ac{EWS} area, with an \ac{RMS} of 0.41\,mas and a maximum of 0.88\,mas, well below the requirement of 3\,mas. The remaining systematic offset, $(\Delta X, \Delta Y) = (0.30, -0.04)$\,mas, is partially caused by the chromatic shift effect discussed in \cref{sc:chromatic-shift}. The right panel shows the residuals after subtracting this offset, averaged over all quadrants, which have an \ac{RMS} of 0.28\,mas. This map averages over all exposures. For individual exposures, the mean astrometric residual computed by the quality control (\cref{fig:dr1-dqc}) is below 3\,mas for $99.8\%$ of DR1 images. 

Examining the right panel in \cref{fig:wide_astro_field_comparison}, we see that the residuals differ between adjacent quadrants. To investigate this, we select all point-like sources in the \ac{EWS} and \ac{EDS} within $\pm11$\,mm of the boundaries between the four quadrants of each \ac{CCD}, and compute their mean $\Delta X$ and $\Delta Y$ in bins of signed distance to the nearest boundary, combining all boundaries of the same orientation. There is a coherent variation in the mean residual profile, with a peak-to-peak amplitude of around 0.2\,mas.

Weak lensing shear measurements are sensitive to the differential distortion, which induces a spurious shear $\gamma$ that can bias the measured galaxy ellipticities. Moreover, the \ac{WCS} solutions are fitted independently for each \ac{VIS} quadrant, and this may lead to larger residuals at the quadrant boundaries and thereby contribute to additional spurious ellipticity. To quantify this, we examine how this residual depends on scale and connect it to the $3\times10^{-5}$ requirement. 

We consider VIS$-$\Gaia\ residuals $(\Delta X, \Delta Y)$ for all point-like sources with $18.5\le\IE<19.8$ in the \ac{EWS} and \ac{EDS} nominal exposures, without the \Gaia quality cuts  ($9\,587\,166$ measurements). We define strips of width $L$ on either side of the vertical and horizontal quadrant boundaries inside each \ac{CCD} and compute the differences between the mean astrometric residuals of all sources on either side, using $\Delta Y$ across vertical and $\Delta X$ across horizontal boundaries. Dividing by the separation $b$ of the mean source positions in the two strips (which accounts for the narrow gap empty of sources at the boundary) provides a finite-difference estimate of the displacement gradient across the quadrant edges. For each orientation, we take the \ac{RMS} of the steps $s_q$ over the 72 quadrant pairs $q$ (the internal boundary of each CCD is crossed by two independently fitted pairs), and obtain the edge shear
\begin{equation}
\gamma_{\rm V,H}(L) = \frac{1}{2\,b}\sqrt{\langle s_q^2\rangle_q - \langle s_q^2\rangle_{q,\rm noise}}\,,
\end{equation}
where the residuals and $b$ are expressed in the same units and the second term is the noise floor.\footnote{The noise floor is computed by repeating the measurement with the same stars with randomly rotated residual vectors. This removes any spatially coherent distortion while keeping the same length for the residual vector of each star and therefore the measurement noise. The median of 60 of these realisations is subtracted in quadrature at each angular scale.} These are combined in quadrature to obtain the total spurious edge ellipticity, $|e| = 2\sqrt{\gamma_{\rm V}^2 + \gamma_{\rm H}^2}$, where $e = 2\gamma$. 

\begin{figure}[htbp]
  \centering
  \includegraphics[width=\hsize]{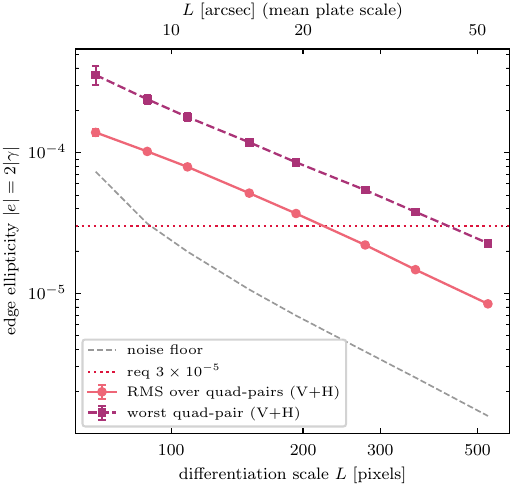}
  \caption{Spurious edge ellipticity as a function of scale $L$ as  measured from the astrometric residuals of \Gaia\ stars in the DR1 \ac{EWS} and \ac{EDS}. The solid line shows the total spurious ellipticity obtained by combining the vertical and horizontal edge components, while the dashed line with markers shows the combination of the largest vertical and horizontal values measured for any quadrant pair. The horizontal dotted line indicates the requirement value of $3\times10^{-5}$ (see text), and the grey dashed line represents the estimated noise floor.}
  \label{fig:edge-ellipticity}
\end{figure}

The total spurious edge ellipticity is shown as the solid line in \cref{fig:edge-ellipticity}. The dashed line shows the combination of the largest vertical and horizontal values found for any quadrant pair, and the dotted line the $3\times10^{-5}$ requirement level. At scales smaller than \ang{;;22}, we exceed this requirement. However, we note that the requirement is defined after averaging within \ang{;4;}, so this comparison at smaller scales is more stringent than the requirement itself. The source of this systematic error will need to be addressed for DR2. 

\subsubsection{\label{sc:AstAccIntExto} Internal and external astrometric residuals}

\begin{figure}[htbp]
    \centering
    \includegraphics[width=1\hsize]{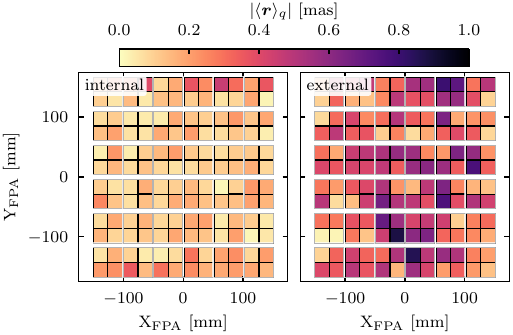}
    \hfill
    \caption{Internal and external astrometric residuals across the VIS focal plane for the DR1 \ac{EWS}. Each of the 144 quadrants is colour-coded by the norm of the mean signed residual vector in the focal-plane frame $|\langle\boldsymbol{r}\rangle_q|$ (\cref{eq:mean}), computed between VIS positions and \Gaia DR3 coordinates (right), and between VIS positions and the mean position of each source over its repeated observations, after subtracting the median offset of each observing sequence (left).}
    \label{fig:int-external-astrom}
\end{figure}

We now examine the internal and external astrometric performance of \ac{VIS} \ac{PF} computed by quadrant over the focal plane for the same $18.5<\IE<19.8$ reference bin as before. As described in \cref{sc:CatCreation}, the grouped catalogue contains all observations of each source, identified by a unique \texttt{group\_id}, together with their cross-matches to the \Gaia\ reference catalogue. Both measurements are shown in \cref{fig:int-external-astrom}. In the left panel, for sources with multiple observations, we compute the difference between each individual VIS-measured position and the mean group position, rotate it into the focal-plane frame with \cref{eq:xy}, and subtract the median difference of each observing sequence (usually four dithered nominal exposures). In the right panel, grouped sources are compared with their matched \Gaia\ DR3 reference source, propagated to the VIS observation epoch and corrected for parallax. In both panels, we report the norm of the mean signed residual vector in a cell $c$,
\begin{equation}
|\langle\boldsymbol{r}\rangle_c| = \sqrt{\langle\Delta X\rangle_c^2 + \langle\Delta Y\rangle_c^2}\,,
\label{eq:mean}
\end{equation}
where $\langle\cdot\rangle_c$ is the mean over all measurements in the cell. In \cref{fig:int-external-astrom} the cells are the 144 quadrants, pooled over the 49 skypatches. \Cref{tab:dr1-internal-external} summarises these measurements and also gives the source-level and patch-level statistics. We find a source-level (quadrant-level) internal single-measurement repeatability (corrected for the contribution of each measurement to its group mean) of $\sigma_{\rm s} = 3.4\,\mathrm{mas}$ ($\sigma_{\rm q} = 0.08\,\mathrm{mas}$) and an external source-level (quadrant-level) dispersion of $\sigma_{\rm s} = 5.6\,\mathrm{mas}$ ($\sigma_{\rm q} = 0.18\,\mathrm{mas}$).

In both cases, we measure a small residual mean norm (\cref{eq:mean}) of  $0.14\,\mathrm{mas}$ (internal) and $0.37\,\mathrm{mas}$ (external) on the quadrant level. For weak lensing, there is no formal requirement on astrometric repeatability and accuracy. 

\begin{figure}[htbp]
    \centering
    \includegraphics[width=1\hsize]{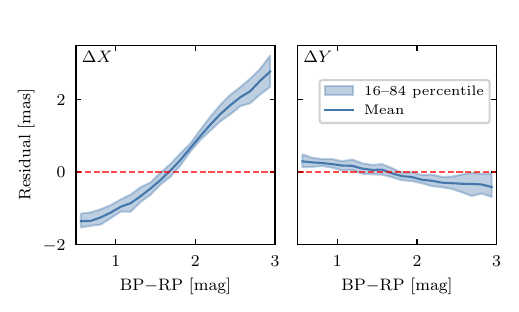}
    \hfill
    \caption{Mean per-source residual in $\Delta X$ (left) and $\Delta Y$ (right) as a function of $G_{\mathrm{BP}}-G_{\mathrm{RP}}$ averaged over all DR1 \ac{EWS} observations. The shaded region shows the $16$th--$84$th percentile region of the per-skypatch means in each colour bin.}
    \label{fig:cshift-xy}
\end{figure}

\subsubsection{\label{sc:chromatic-shift}The chromatic shift effect}

The `chromatic shift' is a colour-dependent displacement of object centroids in VIS images. It arises from a combination of chromatic tip-tilt \ac{WFE} from the dichroic mirror and a wavelength-dependent lateral shift of absorbed photons within the \acp{CCD} because the beam falls on the detectors at a non-normal angle of incidence that varies across the focal plane (Miller et al., in prep.). Using the VIS$-$\Gaia residuals computed in \cref{eq:xy} we can characterise this displacement in two ways. First, we bin sources by \Gaia\ $G_{\mathrm{BP}} - G_{\mathrm{RP}}$ colour and compute the mean per-source $\Delta X$ and $\Delta Y$ residuals in detector coordinates, averaged over all DR1 \ac{EWS} observations. This is plotted in \cref{fig:cshift-xy}, which shows that redder sources are shifted primarily in $X_{\rm FPA}$. 

\begin{figure}[htbp]
  \centering
  \includegraphics[width=\hsize]{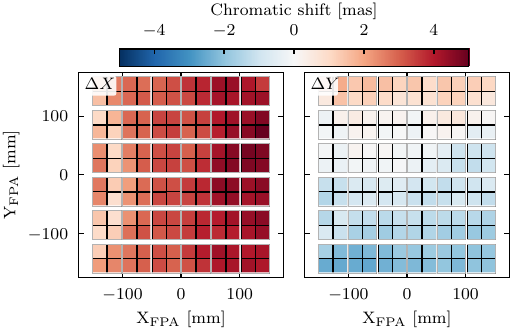}
  \caption{%
    Per-quadrant chromatic shift (red minus blue) across the VIS \ac{FPA} for the DR1 \ac{EWS}, shown separately for the $\Delta X$ (left) and $\Delta Y$ (right) components. The diverging colour scale is centred on zero and shared between panels.%
  }
  \label{fig:chromatic-shift-global}
\end{figure}

Next, we investigate the dependence of chromatic shift on position in the \ac{FPA} by dividing our $18.5<\IE<19.8$ reference sample into two arbitrary colour bins: a red bin with $2.4 < G_{\mathrm{BP}} - G_{\mathrm{RP}} < 2.6$ and a blue bin with $0.65 < G_{\mathrm{BP}} - G_{\mathrm{RP}} < 0.85$. In each bin, the mean VIS$-$\Gaia\ position residual is averaged first per observation and per quadrant, then averaged across observations. The chromatic shift of each quadrant is the red-minus-blue difference of these quadrant-averaged means, measured separately for the $\Delta X$ and $\Delta Y$ components. The resulting map is shown in \cref{fig:chromatic-shift-global}.

One important consequence of this effect is that, as it is focal-plane-position dependent, observations of the same object at different \acp{PA} will be shifted  slightly. This can be seen clearly when (for example) comparing the dependence of astrometric residuals as a function of \ac{PA} for objects in the self-calibration field, where there are observations at many \acp{PA}. Consequently, applications in \ac{DR1} requiring sub-mas precision at the calibrated frame level will require a correction for the chromatic shift effect (Appendix~\ref{sc:cshift-table}). 

\subsection{\label{sc:PhotAccInfo}Photometric accuracy of the merged DR1 catalogues}

We compared the photometry of the $18.5<\IE<19.8$ \ac{EWS} reference sample  with matched \Gaia sources as well as repeat observations of the same source, considering a hierarchy of increasing spatial scales: individual sources; quadrants; \acp{CCD}; the \ac{FPA}; skypatches; and the full DR1 set of \ac{EWS} skypatches. We used the 13-pixel-diameter fixed-aperture photometry in the VIS catalogues, from which we subtracted the supplied aperture correction, ${\rm \texttt{APERCORR}}=0.0655$, derived from the mean difference between stellar fluxes measured in 13- and 50-pixel diameter apertures. 

\begin{figure}[htbp]
    \centering
    \includegraphics[width=1\hsize]{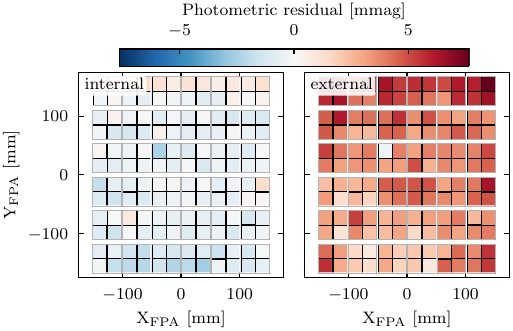}
\caption{Internal and external photometric residuals across the VIS focal plane for the DR1 \ac{EWS}. Each of the 144 quadrants is colour-coded by its mean photometric residual on a diverging colour scale centred on zero. \emph{Left}: internal residuals, the difference between each observation of a source and its mean magnitude (\cref{sc:PhotAccInfo}); the quadrant means range from $-2.7$ to $+1.5\,\mathrm{mmag}$ and the per-quadrant single-measurement scatter from $10.5$ to $12.5\,\mathrm{mmag}$. \emph{Right}: external residuals, VIS$-$\Gaia DR3; the quadrant means range from $-0.3$ to $+7.6\,\mathrm{mmag}$ and the per-quadrant scatter from $24$ to $26\,\mathrm{mmag}$.}

\label{fig:phot-internal-external-residuals}
\end{figure}

Each matched source has \Gaia $G$, $G_{\mathrm{BP}}$ and $ G_{\mathrm{RP}}$ magnitudes. These magnitudes are transformed to $\IE$ as described in Appendix~\ref{sc:gaia-ie}. 

Internal photometric residuals are given by the magnitude difference between each source observation and the mean magnitude of all its observations, $\IE-\langle\IE\rangle$,  after subtracting the median residual of each observing sequence (usually four dithered nominal exposures). For external residuals we compute $\IE-G'$, where $G'$ is the \Gaia magnitude transformed to the \IE band.
Before computing these statistics, we reject measurements in each skypatch whose external residual deviates from the patch median by more than three times the normalised median absolute deviation; this removes $4.5\%$ of the measurements, which are excluded from both the internal and external statistics. Around half of the rejected measurements belong to sources that are rejected in every exposure, more than half of which are classified by \Gaia as quasars or variables, for which the stellar \Gaia-to-\IE transformation is not valid; the remainder are isolated discrepant measurements of otherwise well-behaved stars. As for the astrometry, each cell is characterised by the mean residual over all measurements in the cell, in this case the signed mean $\langle\Delta m\rangle_c$. The quantity $\sigma_{\rm s}$ is the repeatability of a single measurement, computed with $n-1$ degrees of freedom for a source observed $n$ times. The small remaining internal mean of $-0.5\,\mathrm{mmag}$ is a consequence of subtracting the sequence medians from residuals with a tail at brighter values.  After applying a $5\,\sigma$ clip to internal residuals it becomes $+0.09\,\mathrm{mmag}$. 

\Cref{fig:phot-internal-external-residuals} shows these quantities averaged at quadrant level. Internally, we find $\sigma_{\rm s} = 11\,\mathrm{mmag}$ and $\sigma_{\rm q} = 0.76\,\mathrm{mmag}$ at the source and quadrant level, respectively, where the latter averages out the shot noise. Externally, we find $\sigma_{\rm s} = 25\,\mathrm{mmag}$ and $\sigma_{\rm q} = 1.4\,\mathrm{mmag}$ and a mean offset of $3.8\,\mathrm{mmag}$ with respect to \Gaia, well below the 5\% requirement on the absolute photometric error (R-CAL-A-VS-5000 in \citealt{DR1-TP026}).

Because each source is observed at several dither patterns, the internal quadrant means provide an excellent test of the accuracy of our illumination correction at quadrant scales. The statistical uncertainty in the means at each quadrant is small ($0.05\,\mathrm{mmag}$) and the internal and external maps are strongly correlated, so the same outlier quadrants are visible in both. The largest internal offsets, up to about $2\,\mathrm{mmag}$ from the mean, are in the outermost CCD rows along $Y_{\rm FPA}$, which are offset by $-0.7$ and $+0.9\,\mathrm{mmag}$ relative to the inner rows.  Even these offsets are well below the requirement that the post-calibration relative photometric error should be less than $1\%$ (around $11\,\mathrm{mmag}$; MRD-WL-020 in \citealt{DR1-TP026}).

\subsection{\label{sc:selfcal}Self-calibration field observations}

The self-calibration field is observed each month as part of regular operations. Each set of observations is known as a `visit' and comprises 60 dithers with nominal and short exposures; 16 additional dithers are repeated to accommodate the blue-grism calibration \citep{DR1-TP026}, giving a total of 76 nominal and 72 short exposures per visit. We now analyse calibrated frame catalogues from these exposures using the same procedure as previously. We construct a merged catalogue of \Gaia-matched VIS measurements from the 16 self-calibration visits obtained between October~2023 and May~2025 (including four visits from before the start of the survey to extend the time range), containing $4\,167\,347$ measurements grouped into $11\,245$ matched sources. Restricting to nominal exposures and selecting sources with $\mathrm{SNR}\geq 10$, $18.5<\IE<19.8$, and detections in at least two visits yields a photometric reference sample of $4352$ stars.

The first two rows of \cref{tab:selfcal-scatter} report the \ac{RMS} visit-to-visit scatter of the reference-ensemble median offsets ($1.5\,\mathrm{mmag}$ over all visits, $0.55\,\mathrm{mmag}$ post-decontamination) and the scatter of the per-visit median VIS$-$\Gaia\ residuals ($1.6\,\mathrm{mmag}$ / $0.90\,\mathrm{mmag}$). The factor of three reduction after decontamination indicates that the applied photometric corrections track the associated throughput variations to about $1\,\mathrm{mmag}$; the same reference sample records a throughput change of $4$--$5\,\mathrm{mmag}$ across the decontamination interval. Because the same self-calibration visits are used to derive the zero-point model, this is not a fully independent test; a more robust test would exclude each visit from the calibration in turn. The wide-survey photometric data in \cref{sc:PhotAccInfo}, which does not enter the calibration, provides such an independent check, with quadrant-level internal and external dispersions of $0.76$ and $1.4\,\mathrm{mmag}$.

As an independent cross-check, we compute residuals between the white-dwarf reference standard J175318+644502, an \textit{HST} secondary flux standard in the self-calibration field (section~7.2 in \citealt{DR1-TP026}), and the nominal-exposure reference sample, finding a scatter of $9.6\,\mathrm{mmag}$ over all visits and $6.6\,\mathrm{mmag}$ post-decontamination (last row of \cref{tab:selfcal-scatter}). Given this substantially larger scatter, we prefer photometric zero-point determinations based on \Gaia-matched stars. These self-calibration uncertainties are consistent with the quadrant-level $\sigma_{\rm q}$ values of $0.76\,\mathrm{mmag}$ (internal) and $1.4\,\mathrm{mmag}$ (external) derived over the entire \ac{EWS}  (\cref{sc:PhotAccInfo}, \cref{tab:dr1-internal-external}).

\begin{table}[htbp]
\caption{Self-calibration photometric scatter for the $4352$-star nominal-exposure reference sample, for all visits and for the visits after the decontamination event.}
\label{tab:selfcal-scatter}
\centering
\begin{tabular}{l S[table-format=1.1] S[table-format=1.2]}
\hline\hline
\noalign{\smallskip}
Quantity & {All} & {Post-dec.} \\
         & {[mmag]} & {[mmag]} \\
\hline
\noalign{\smallskip}
Reference-ensemble visit scatter  & 1.5 & 0.55 \\
Per-visit VIS$-$\Gaia\ residual   & 1.6 & 0.90 \\
White-dwarf vs.\ reference sample & 9.6 & 6.6  \\
\noalign{\smallskip}
\hline
\end{tabular}
\end{table}

\subsection{\label{sc:deep}Deep field observations}

The \ac{EDS} \citep{EuclidSkyOverview} contains three deep fields (Euclid Deep Field North, South, and Fornax) comprising 23 skypatches in \ac{DR1} which are repeatedly observed throughout the survey. In \ac{DR1}, some deep skypatches contain up to 52 separate visits. We analysed the calibrated frame catalogues from all \ac{DR1} deep fields. We find that photometric and astrometric accuracy is very similar to the figures reported here for the wide survey. Over the deep skypatches, the quadrant-level internal and external astrometric dispersions are $\sigma_{\rm q} = 0.07$ and $0.18\,\mathrm{mas}$, and the photometric dispersions are $0.50$ and $1.2\,\mathrm{mmag}$ (with $0.08$, $0.18\,\mathrm{mas}$ and $0.76$, $1.4\,\mathrm{mmag}$ for the wide survey; \cref{tab:dr1-internal-external}). The external source-level dispersions, $5.7\,\mathrm{mas}$ and $24\,\mathrm{mmag}$, are also almost the same.

\subsection{\label{sc:iq}Image quality}

\begin{figure}[htbp]
  \centering
  \includegraphics[width=0.8\hsize]{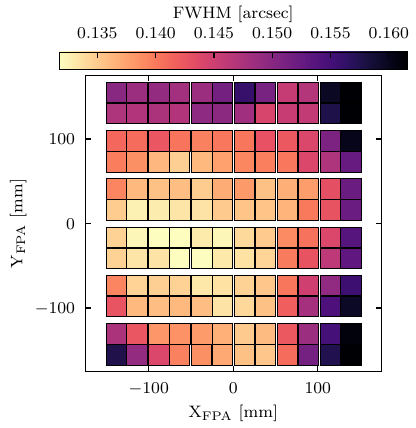}
  \caption{Median PSF \texttt{FWHM\_IMAGE} across the VIS focal plane for \Gaia-matched sources in the DR1 \ac{EWS} reference sample. The colour bar limits cover 2nd--98th percentiles of the distribution. The focal plane median FWHM is $0\farcs140$, with a $16$--$84$ percentile range across quadrants of  $0\farcs135$--$0\farcs150$; edge quadrants reach $\sim 0\farcs176$.}
  \label{fig:dr1-psf-fpa-fwhm}
\end{figure}

\Cref{fig:dr1-psf-fpa-fwhm} shows the median \ac{FWHM} of sources in our \ac{EWS} stellar reference catalogue in each quadrant over the focal plane. The \ac{FWHM} measurement is SExtractor's \texttt{FWHM\_IMAGE} which assumes a Gaussian profile. Even the extreme quadrant values are below the requirement that the \ac{PSF} \ac{FWHM} should be smaller than \ang{;;0.18} (MRD-WL-002 in \citealt{DR1-TP026}). 

\begin{figure}[htbp]
  \centering
  \includegraphics[width=\hsize]{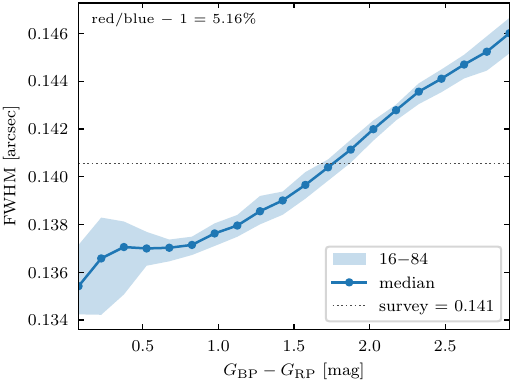}
  \caption{Median stellar \ac{FWHM} as a function of \Gaia\
  $G_{\rm BP}-G_{\rm RP}$ colour for the same sample, averaged over all
  DR1 \ac{EWS} exposures. The shaded band shows the $16$--$84$ percentile
  patch-to-patch range and the dotted horizontal line marks the survey
  median. The label gives the ratio of the median \ac{FWHM} of stars with $G_{\rm BP}-G_{\rm RP}>2$ to that of stars with $G_{\rm BP}-G_{\rm RP}<1$, minus one.}
  \label{fig:dr1-psf-fwhm-vs-bprp}
\end{figure}

The broad bandpass of VIS (see appendix E in \citetalias{Q1-TP002}) means that for a given source, the measured \ac{FWHM} in VIS depends on intrinsic colour. \Cref{fig:dr1-psf-fwhm-vs-bprp} shows the clear dependence of \ac{FWHM} as a function of  $G_{\mathrm{BP}}-G_{\mathrm{RP}}$. For a sample of stars split at $G_{\mathrm{BP}}-G_{\mathrm{RP}}=1.5$, red stars have a median \ac{FWHM} $3.3\%$ larger than the blue stars in the same quadrant (mean over all quadrants). We also found that this colour-dependent effect varies by $0.5$\% (\ac{RMS}) over all quadrants in the focal plane.

Interestingly, the bump visible at $0.30 < G_{\mathrm{BP}}-G_{\mathrm{RP}} < 0.45$ (and which is even clearer in \texttt{FLUX\_RADIUS}) is a consequence of \ac{QSO} contamination. In this colour range, $56\%$ of the sources have proper motions consistent with zero, compared with $3$--$5\%$ for those with $G_{\mathrm{BP}}-G_{\mathrm{RP}} > 0.75$.

\subsection{\label{sc:bfe-nl}Testing for residual brighter-fatter or non-linearity effects}

\Ac{NL} means that for brighter sources, the number of \acp{ADU} read out from the detector is a non-linear function of the number of incident photons. The \ac{BFE} is a consequence of the differing charge distributions between  brighter and fainter sources. It leads to a \ac{PSF} which is more extended for brighter sources and narrower for fainter sources.

In DR1, both \ac{BFE} and \ac{NL} are corrected using techniques outlined in sections 3.9 and 3.6, respectively, of \citetalias{Q1-TP002}. In general, these effects are interrelated and it can be difficult to separate them. However, we attempted to see if residual uncorrected \ac{BFE} or \ac{NL} could be detected in our reference star catalogue by comparing observations of the same source in short and nominal exposures. 

\begin{figure}[htbp]
  \centering
\includegraphics[width=\hsize]{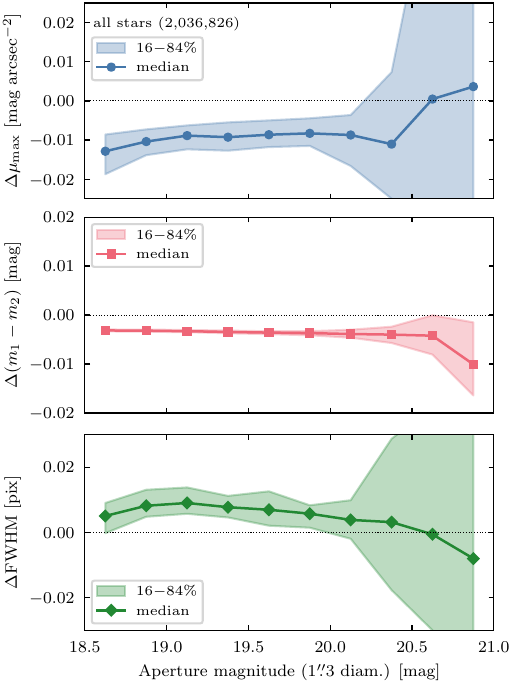}
  \caption{Nominal-minus-short differences for \Gaia-matched stars in
  nominal and short exposures, for $1\farcs3$-diameter aperture magnitudes in the overlap interval $[18.5, 21.0]$. For each source and quantity $P$, $\Delta P = \langle P \rangle_{\mathrm{nominal}} - \langle P
  \rangle_{\mathrm{short}}$, where $\langle P \rangle$ denotes the median of all observations of the source in nominal and short exposures, respectively.
  Panels show, from top to bottom: peak surface brightness $\Delta\mu_{\mathrm{max}}$; the curve-of-growth metric  $\Delta(m_1 - m_2)$, where $m_1$ and $m_2$ are the \texttt{MAG\_APER\_1} and \texttt{MAG\_APER\_2} magnitudes in apertures of $0\farcs7$ and $1\farcs3$ diameter, respectively; and image \ac{FWHM}. Solid lines are the  median across all 50 \ac{EWS} skypatches ($2\,036\,826$ matched stars), while the shaded bands show the $16$--$84$ percentile range of per-patch medians in each magnitude bin.}
  \label{fig:dr1-bfe-limits}
\end{figure}

This is illustrated in \cref{fig:dr1-bfe-limits}. For each unique source, we consider the difference of the following quantities between nominal and short exposures as a function of aperture magnitude: the peak surface brightness; the difference between the first and second aperture magnitudes; and lastly, the \ac{FWHM}. When there are multiple observations of the same object, per-exposure-type medians were computed. Within our measurement uncertainties, we detect no dependence of these quantities on magnitude, and therefore no evidence of any residual \ac{BFE} or \ac{NL} in our data. However, small constant offsets are visible in each panel. In the top panel, we interpret a slightly negative nominal-minus-short peak surface brightness as consistent with a slight overcorrection of the BFE. The second panel shows a constant magnitude offset with aperture magnitude. This was already noticed in \citetalias{Q1-TP002}; it is consistent with the uncorrected shutter effect. Finally, the slightly positive FWHM difference we report also depends on $X_{\rm{FPA}}$ in an equivalent manner to the shutter correction. 

\subsection{\label{sc:eudf}Testing the depth-scaling relationship in the Euclid Ultra-Deep Field}

All tests presented until now have been based on statistics of bright, unsaturated stars in single \ac{VIS} exposures. As an additional quality control check, we investigate the properties of stacked \ac{VIS} data. The \ac{EUDF} will be one of the deepest optical images ever made, and comprises coadded \ac{VIS} and \ac{NISP} exposures taken in the self-calibration field. In \ac{DR1}, there are more than 3000 VIS exposures in this region. We can use the \ac{VIS} \ac{DR1} \ac{EUDF} data to investigate the scaling relationship between exposure time and depth and to investigate if there are any residual sources of systematic error or noise.  

A complete description of the \ac{EUDF} processing is given in a forthcoming paper (McCracken et al., in prep.). The self-calibration dither pattern (figure~3 in \citealt{DR1-TP026}) means that after coaddition each region of the field has a different effective exposure time and as a consequence differing depths. In the \ac{EUDF}, all VIS exposures have been coadded to a pixel scale of 50\,mas in a series of tiles using a custom pipeline based on the \texttt{Drizzle} \citep{2002PASP..114..144F} algorithm.  Over a grid in each tile, we compute the depth by measuring the variance of the sky background in empty $0\farcs5$ apertures on 100\,mas block-summed mosaics after masking object pixels. By comparing the depth at each location with the total exposure time that contributed to that measurement, we can derive the exposure time--depth relationship. We also exclude any aperture that falls on the extended wings or artefacts of a bright star not captured by our object mask, built using \Gaia DR3 stars. 

\begin{figure}[htbp]
  \centering
  \includegraphics[width=\hsize]{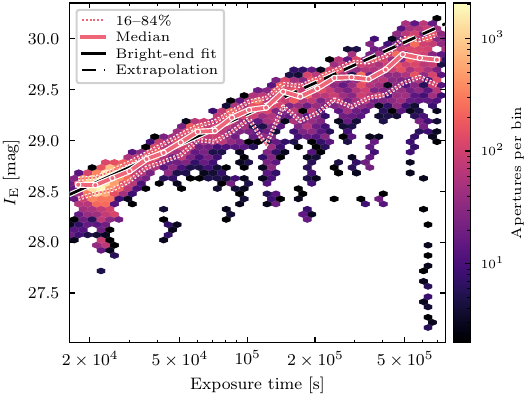}
  \caption{$5\,\sigma$ \IE limiting magnitude as a function of exposure time in the \ac{EUDF}. The two-dimensional histogram shows the individual depth measurements, made in empty $0\farcs5$ apertures after masking bright stars and artefacts, and is shaded by the number of apertures in each bin. The solid red line is the median depth in bins of exposure time, and the dotted red lines the 16th and 84th percentiles. The dashed black line is the extrapolation of a fit made at shorter exposure times (solid line). We attribute the 0.2\,mag deviation of the red line from the extrapolated relation to the presence of residual large-scale object flux.}
  \label{fig:eudf-depth-exp}
\end{figure}

This is shown in \cref{fig:eudf-depth-exp}, where the red line shows the median $5\,\sigma$ depth as a function of exposure time and the solid black line a straight-line fit over  $1.5\times10^4 < t < 8\times10^4$\,s. The extrapolation of this fit to longer exposures is shown as the dashed line. The slope gradually flattens, falling 0.2 mag below the extrapolated relation in the deepest regions. To investigate the source of this noise floor, we experimented with changing the object masking threshold. If there was an unaccounted systematic linked to residual flat-field errors, or a detector residual, it should not depend on masking threshold.  

\begin{figure}[htbp]
  \centering
  \includegraphics[width=\hsize]{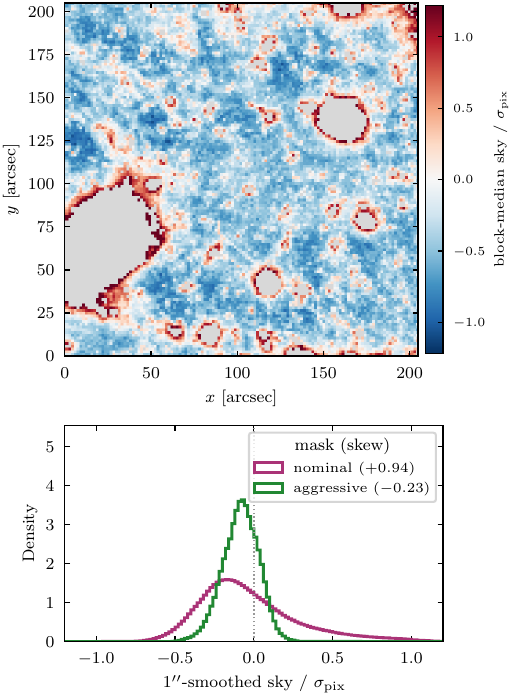}
  \caption{\emph{Top panel}: Median of the source-masked sky in 1\farcs6 cells in the deepest \ac{EUDF} tile in per-pixel noise units. The red haloes around the grey-masked sources are unmasked low-surface-brightness wings of galaxies. We note that there are many unmasked sources. The background level is slightly negative due to oversubtraction by \texttt{NoiseChisel}. \emph{Bottom panel}: distribution of sky pixels smoothed  at 1\arcsec scales with two different masks. The nominal masking curve (purple) shows clearly the presence of objects in the extended tail, whereas the aggressive masking curve (green) removes them. }
  \label{fig:eudf-noise-floor}
\end{figure}

The top panel of \cref{fig:eudf-noise-floor} shows the median of the source-masked sky in the deepest \ac{EUDF} tile together with the distribution of smoothed sky pixels (bottom panel). Nominal masking results in pixels clearly polluted by object flux. A correlation analysis shows that this extra light is present at all scales. Lowering the masking threshold removes it, but leaves almost no sky to measure. The excess is therefore attached to the detected sources, and is consistent with the change in slope in the exposure time--depth relation arising from the diffuse light of faint galaxies. We cannot, however, exclude a contribution from unmasked ghosts or  residuals in the background subtraction. 

\section{\label{sc:caveats} Known caveats}
\subsection{\label{sc:psf} PSF estimation}

During DR1 validation, we found that photometric quantities estimated from \ac{PSF} fitting (e.g., \texttt{MAG\_MODEL}) using the \ac{VIS} \ac{DR1} \acp{PSF} had a larger-than-expected scatter. The origin of this scatter is an inconsistent \ac{PSF} normalisation between quadrants, which is caused by over-fitting of each quadrant's \ac{PSF} by using a second-order polynomial to model the variation of the \ac{PSF} with position on the detector. In some quadrants, there are not enough stars to determine the PSF accurately. Therefore, the \texttt{MAG\_MODEL} quantities provided in the calibrated frame catalogues should be treated with caution. We note that PSF-fitted photometry is not used inside \ac{VIS} \ac{PF} and so this \ac{PSF} inaccuracy does not affect any VIS data products beyond the calibrated frame catalogues. In preparation for DR2, we have generated a new PSF fitted with a first-order polynomial, greatly reducing the quadrant-to-quadrant normalisation scatter.

\subsection{\label{sc:imagart} Uncorrected image artefacts}

Several minor artefacts are still present in VIS images. Very bright ($\IE\sim10$) stars can produce bleeding along the serial register. This creates bands in the calibrated exposures and in \ac{MER PF} mosaics. This effect will be corrected in \ac{DR2}. A related effect, readout node  persistence, can smear bright pixels along the serial readout direction. Candidate algorithms are being investigated to correct this effect in \ac{DR2}. In deep images, some unmasked ghosts are still present (visible sometimes as a small diffuse object in the images), and one potential solution involves changing the mask threshold for the parent ghost star. These unmasked ghosts can produce spurious detections in the \ac{MER PF} catalogues; improved masking will be delivered as part of \ac{DR2}. 

\subsection{\label{sc:astrpatch3940}Anomalously high astrometric distortions in skypatch 40}

While validating the DR1 production, it was found that the focal plane residual distortion increased sharply between 15 and 18 June 2024, during observations of skypatch 40 (\ac {EWS}) and skypatch 39 (\ac{EDS}). When considering the difference between the two successive distortion models used around this period, it appeared that the VIS focal plane underwent a significant deformation or warping along its top-right to lower-left diagonal. The origin of this deformation is unclear, but it occurred shortly after the second ice decontamination (on 8 June 2024) and may have been caused by temperature variations induced by the decontamination. Both skypatches were processed with the same distortion model, but only skypatch 40 shows a measurable anomaly: in these exposures the largest per-quadrant mean residual has a median of 23\,mas, compared with 9\,mas (and at most 17\,mas) in the other skypatches; skypatch 39 is within the nominal range. The mean residual over the focal plane remains below the requirement. We therefore exclude skypatch 40 from the astrometric analysis in \cref{sc:AstAcc}. Fortunately, skypatch 40 was immediately followed by a self-calibration visit (on 19 June 2024), from which an updated distortion model was derived and used. No other significant focal plane deformation was detected in \ac{DR1}.

\section{\label{sc:Conc} Conclusions}
We have presented the data products and processing algorithms used to produce VIS DR1 data products. These data products meet the stated requirements, with astrometric and photometric precision quantified at the source, patch, and quadrant level. We characterised the chromatic shift, which may need to be corrected in future data releases. 

Although the \ac{RMS} of the quadrant-averaged astrometric residuals is 0.41\,mas, and the mean residual per exposure is below the $3\,\mathrm{mas}$ requirement for $99.8\%$ of DR1 images, a more detailed investigation shows that on scales below \ang{;;22} the spurious edge ellipticity exceeds the $3\times10^{-5}$ level, although the requirement itself is defined after averaging within \ang{;4;}. These  distortion residuals are sufficient for DR1 but will need to be improved for DR2. No evidence of residual brighter-fatter-effect or non-linearity is seen in our data products. The noise floor in the deepest VIS images appears to be set by the extended diffuse light of faint galaxies. 

At the time of writing (November 2026) preparations for DR2 are well underway. The next data release will provide VIS data over more than three years of observations, and is expected to cover approximately $5800\,\mathrm{\deg^2}$ and contain more than  120\,000 VIS calibrated frames. The uncompressed data volume including associated products is expected to be more than 1.5\,PB. Such a large data volume will require a rigorous approach to validation and quality control, but we are confident in the foundations we have laid down here in DR1.

\begin{acknowledgements}
This work was made possible by utilising the CANDIDE cluster at the Institut d'Astrophysique de Paris. The cluster was funded through grants from the PNCG, CNES, DIM-ACAV, the Euclid Consortium, and the Danish National Research Foundation Cosmic Dawn Center (DNRF140). It is maintained by Stéphane Rouberol. We also acknowledge support from the CNRS/IN2P3 Computing Centre (Lyon, France) for providing computing and data-processing resources needed for this work. Generative artificial intelligence tools were used to assist with the analysis presented in this paper. The authors reviewed and verified the resulting code and outputs and take full responsibility for the analysis, results, and content of this paper.
  \AckDatalabs
  \AckEC
  \AckDRone
\end{acknowledgements}

%
%

\bibliography{Euclid,manual-refs}

%
%

\begin{appendix}
\section{\label{sc:gaia-ie}Transformation from \texorpdfstring{\Gaia}{Gaia} to \texorpdfstring{\IE}{IE} magnitudes}

Throughout this paper, predicted \IE\ magnitudes of \Gaia\ stars are computed from the \Gaia\ $G$, $G_{\mathrm{BP}}$, and $G_{\mathrm{RP}}$ magnitudes using a transformation stored in the \Euclid mission database. For completeness, we reproduce this transformation here.

The \Gaia\ magnitudes are first converted from the Vega to the AB system by adding $0.1136$, $0.0155$, and $0.3561$\,mag to $G$, $G_{\mathrm{BP}}$, and $G_{\mathrm{RP}}$, respectively. With the AB colour
\begin{equation}
C = G_{\mathrm{BP}} - G_{\mathrm{RP}} - 0.3406\,,
\label{eq:gaia-ie-colour}
\end{equation}
the predicted magnitude is
\begin{equation}
\IE = G + 0.1136 - f(C)\,,
\label{eq:gaia-ie}
\end{equation}
where $f(C)$ is the tabulated relation given in \cref{tab:gaia-ie}, valid for $0.079 < C < 2.921$. Sources outside this colour range are not used in the photometric analysis. Linear interpolation in \cref{tab:gaia-ie} reproduces the original relation, sampled at 1000 points, to better than 2\,mmag.

\begin{table}[htbp]
\caption{Tabulated \Gaia-to-\IE\ relation $f(C)$ of \cref{eq:gaia-ie}, as a function of the AB colour $C$ (\cref{eq:gaia-ie-colour}).}
\label{tab:gaia-ie}
\centering
{\small\setlength{\tabcolsep}{4pt}%
\begin{tabular}{cc@{\hspace{12pt}}cc@{\hspace{12pt}}cc}
\hline\hline
\noalign{\vskip 1pt}
$C$ & $f(C)$ & $C$ & $f(C)$ & $C$ & $f(C)$ \\
{[mag]} & {[mag]} & {[mag]} & {[mag]} & {[mag]} & {[mag]} \\
\hline
0.079 & 0.0432 & 1.100 & 0.3706 & 2.100 & 0.5417 \\
0.200 & 0.0990 & 1.200 & 0.3956 & 2.200 & 0.5496 \\
0.300 & 0.1382 & 1.300 & 0.4195 & 2.300 & 0.5570 \\
0.400 & 0.1658 & 1.400 & 0.4381 & 2.400 & 0.5676 \\
0.500 & 0.1963 & 1.500 & 0.4537 & 2.500 & 0.5731 \\
0.600 & 0.2328 & 1.600 & 0.4723 & 2.600 & 0.5762 \\
0.700 & 0.2621 & 1.700 & 0.4907 & 2.700 & 0.5844 \\
0.800 & 0.2899 & 1.800 & 0.5079 & 2.800 & 0.5969 \\
0.900 & 0.3179 & 1.900 & 0.5208 & 2.900 & 0.6139 \\
1.000 & 0.3432 & 2.000 & 0.5318 & 2.921 & 0.6181 \\
\hline
\end{tabular}}
\end{table}

\section{\label{sc:cshift-table}Chromatic shift as a function of colour}

\Cref{tab:cshift} lists the chromatic shift of \cref{fig:cshift-xy} in bins of $G_{\mathrm{BP}}-G_{\mathrm{RP}}$ and of $G-G_{\mathrm{RP}}$. For red stars ($G_{\mathrm{BP}}-G_{\mathrm{RP}}>1.75$), the parallax-corrected $\Delta X$ at fixed $G_{\mathrm{BP}}-G_{\mathrm{RP}}$ still increases with parallax, by up to 1.4\,mas, whereas at fixed $G-G_{\mathrm{RP}}$ it does not; for these stars $G-G_{\mathrm{RP}}$ is the better predictor of the shift. The astrometric solution of each exposure is fitted to \Gaia\ stars of all colours, so the zero point of \cref{tab:cshift} is set by their mean colour ($G_{\mathrm{BP}}-G_{\mathrm{RP}}\approx1.7$, $G-G_{\mathrm{RP}}\approx0.9$) and varies by 0.1--0.3\,mas between skypatches; only differences between colours are physically meaningful. To correct the position of a star in a single exposure, $(\Delta X, \Delta Y)$ at its colour is subtracted in the focal-plane frame and rotated to the sky with the position angle of that exposure, inverting \cref{eq:xy}.

\begin{table}[htbp]
\caption{Chromatic shift: mean VIS$-$\Gaia\ DR3 astrometric residual along $X_{\rm FPA}$ ($\Delta X$) and $Y_{\rm FPA}$ ($\Delta Y$) after the parallax correction, in bins of $G_{\mathrm{BP}}-G_{\mathrm{RP}}$ and of $G-G_{\mathrm{RP}}$ (\Gaia\ Vega magnitudes), for the \ac{EWS} reference sample of \cref{fig:cshift-xy}; values are relative to the mean colour of the \Gaia\ stars used for the astrometric calibration (see Appendix~\ref{sc:cshift-table}).}
\label{tab:cshift}
\centering
{\small\setlength{\tabcolsep}{4pt}%
\begin{tabular}{crr@{\hspace{12pt}}crr}
\hline\hline
\noalign{\vskip 1pt}
$G_{\mathrm{BP}}-G_{\mathrm{RP}}$ & $\Delta X$ & $\Delta Y$ & $G-G_{\mathrm{RP}}$ & $\Delta X$ & $\Delta Y$ \\
{[mag]} & {[mas]} & {[mas]} & {[mag]} & {[mas]} & {[mas]} \\
\hline
0.50--0.75 & $-$1.37 & 0.29 & 0.30--0.40 & $-$1.39 & 0.30 \\
0.75--1.00 & $-$1.21 & 0.25 & 0.40--0.50 & $-$1.30 & 0.27 \\
1.00--1.25 & $-$0.92 & 0.18 & 0.50--0.60 & $-$1.09 & 0.22 \\
1.25--1.50 & $-$0.57 & 0.09 & 0.60--0.70 & $-$0.85 & 0.16 \\
1.50--1.75 & $-$0.08 & 0.01 & 0.70--0.80 & $-$0.53 & 0.11 \\
1.75--2.00 & 0.50 & $-$0.12 & 0.80--0.90 & $-$0.10 & 0.00 \\
2.00--2.25 & 1.15 & $-$0.22 & 0.90--1.00 & 0.42 & $-$0.11 \\
2.25--2.50 & 1.70 & $-$0.30 & 1.00--1.10 & 1.07 & $-$0.23 \\
2.50--2.75 & 2.12 & $-$0.32 & 1.10--1.20 & 1.77 & $-$0.30 \\
2.75--3.00 & 2.56 & $-$0.37 & 1.20--1.30 & 2.39 & $-$0.36 \\
\dots & \dots & \dots & 1.30--1.40 & 3.08 & $-$0.48 \\
\hline
\end{tabular}}
\tablefoot{Residuals are VIS minus \Gaia\ DR3 positions propagated to the epoch of observation, for sources with $18.5<\IE<19.8$ in nominal \ac{EWS} exposures (skypatch~40 excluded), rotated into the focal-plane frame (\cref{eq:xy}) after subtracting the parallactic displacement (\cref{eq:plx}). In each colour bin we average over sources per quadrant, over quadrants per exposure, and over exposures per skypatch; the values are the means over the 49 skypatches, with standard errors below 0.09\,mas.}
\end{table}

\section{\label{sc:shutter-time} VIS shutter opening time}

Science cases involving transients or fast-moving objects will require the exact (central) time of VIS integration for every exposure. The \texttt{DATE-OBS} header item in the VIS FITS file is described as \fnurl{`UTC Date of observation (start)'}{https://euclid.esac.esa.int/dr/dr1/dpdd}, but there is a varying offset from this time to when the VIS shutter is fully open. While this information is available in the \ac{HKTM}, it is not included in the FITS file header.

\texttt{DATE-OBS} is defined as the time of the beginning of a sequence of events within the VIS instrument. The shutter opening actuation is executed after flushing and surface inversion of the detectors, and the full opening of the VIS \ac{RSU} is typically offset from \texttt{DATE-OBS} by $\sim7.7$\,s (prior to 2024-01-26 00:35:00 UTC, $\sim9.6$\,s), with a spread of $\sim0.1\,$s.

The exact date-time for every event is available in the HKTM (Level\,1 \texttt{DpdHKTMProduct}). For any open-shutter exposure, the VIS instrument records the following four events: \texttt{VIS-START-OPEN-SHUTTER}, \texttt{VIS-SHUTTER-OPEN} (RSU is fully open), \texttt{VIS-START-CLOSE-SHUTTER}, \texttt{VIS-CLOSE-SHUTTER} (RSU is fully closed). These event timestamps show that shutter motion durations are constant at 2.739\,s, for both opening and closing. Hence, the duration of shutter opening should be \texttt{EXPTIME} $+ (2\times2.739)$\,s, but the actual duration of shutter opening, (\texttt{VIS-CLOSE-SHUTTER}$\,-\,$\texttt{VIS-START-OPEN-SHUTTER}), can be off by up to 10\,ms excess due to messaging overhead.

To obtain the true start and end of the integration time, one needs to know when the shutter leaf intersects the beam that covers the VIS focal plane. We estimate that the shutter begins opening the beam 0.17\,s after beginning its motion, and that the beam is fully open after 2.20\,s. Note that the exposure time distribution across the VIS detector arrays has an (approximately linear) gradient due to how the VIS shutter operates; see figure~7 of \cite{DR1-TP026}.

\end{appendix}

\label{LastPage}
\end{document}